\documentclass[preprint,12pt]{elsarticle}
\newcommand{\dslash}[1]{\not\!#1}

\usepackage{graphicx}
\usepackage{wrapft}
\usepackage{float}
\usepackage[tight,TABBOTCAP]{subfigure}
\usepackage{comment}
\usepackage{multirow}

\usepackage[normalem]{ulem}  
\usepackage[dvips]{color} 

\renewcommand\sout{\bgroup \color{red} \ULdepth=-.5ex \ULset}

\usepackage{amssymb}

\journal{Nuclear Physics A}

\begin{document}

\begin{frontmatter}



\title{The $\Lambda^* N$ interaction and two-body bound state \\
based on chiral dynamics}


\author{Toshitaka Uchino, Tetsuo Hyodo, Makoto Oka}

\address{Department of Physics, Tokyo Institute of Technology,\\
Tokyo 152-8551, Japan}

\begin{abstract}
The interaction of $\Lambda^*=\Lambda(1405)$ with a nucleon is studied from the viewpoint of chiral dynamics.
We construct the coordinate space $\Lambda^* N$ potential in the meson-exchange picture, which serves as a fundamental ingredient for the study of the few-body nuclear systems with a $\Lambda^*$, the $\Lambda^*$-hypernuclei.
The coupling constants concerning $\Lambda^*$ are determined based on the chiral unitary model picture for the meson-baryon scattering where $\Lambda^*$ is described as a superposition of two resonance poles.
Solving the coupled-channel two-body $\Lambda^*N$ system, we find the higher energy $\Lambda^*N$ state develops an $s$-wave quasi-bound state slightly below the threshold in the total spin $S=0$ channel, which acquires a finite width through the coupling to the lower energy $\Lambda^*N$ channel.
We show important roles of the $\bar{K}$ exchange contribution to the $\Lambda^*N$ potential.
\end{abstract}


\end{frontmatter}







\section{Introduction}
\label{sec:intro}

One of the most interesting topics of hadron/nuclear physics is possible existence of the $\bar{K}$ bound state in nuclei.
It has been pointed out that the $\bar{K}$-nucleon $s$-wave interaction in isospin $I=0$ channel is strongly attractive and the negative parity hyperon, $\Lambda(1405)=\Lambda^*$ may be described as a $\bar{K}N$ quasi-bound state appearing as a resonance in the $\pi\Sigma$ continuum~\cite{Dalitz:1960du,Dalitz:1967fp}.
The phenomenological interaction in the earlier works is later identified as the leading order term of the $SU(3)\times SU(3)$ chiral perturbation theory, and nonperturbative coupled-channel approach leads to dynamical generation of $\Lambda^*$\cite{Kaiser:1995eg,Kaiser:1995cy,Oset:1998it,Oller:2000fj,Lutz:2001yb,Hyodo:2011ur}.
The chiral $\bar{K}N$ interaction is also a driving force of kaon condensation~\cite{Kaplan:1986yq,Brown:1993yv} when the antikaons are put in the dense nuclear medium.

The strong attraction in the $\bar{K}N$ channel also has an interesting consequence in finite nucleus.
In 2002, it was suggested that $\bar{K}$ can be strongly bound in nuclei so that the corresponding mesonic decay modes are kinematically forbidden and the $\bar{K}$ nucleus may become a narrow state\cite{Akaishi:2002bg}.
Following this, several experimental searches for the bound $\bar{K}$ nucleus were performed, for instance, by KEK E471\cite{Suzuki:2004ep} and E549\cite{Sato:2007sb}, FINUDA at DA$\Phi$NE\cite{Agnello:2005qj}, reanalysis of DISTO experiment\cite{Yamazaki:2008hm,Kienle:2011mi} and so on. 
Some structure was found in the $\Lambda N$ mass spectrum, but the extracted values of the mass and width do not converge quantitatively. 
In addition, it is not clear experimentally that the observed peak structure is caused by the kaon bound state. 
To clarify the experimental situation, the comprehensive analyses will be performed by the E15 experiment at the J-PARC and AMADEUS at DA$\Phi$NE.
Meanwhile, theoretical analyses by rigorous few body calculation for the $\bar{K}NN$ system were done by several groups\cite{Yamazaki:2007cs,Shevchenko:2006xy,Shevchenko:2007zz,Ikeda:2007nz,Ikeda:2008ub,Ikeda:2010tk,Dote:2008in,Dote:2008hw}.
Although these efforts have revealed that there is a quasi-bound state with broad width below the $\bar{K}NN$ threshold, quantitative estimation of the mass and width of the state largely deviates from each other, and the mechanism of the binding is not yet well understood. 
It should be emphasized that the possible experimental signals were found in the $\Lambda N$ spectrum, which has not so far been taken into account explicitly in the theoretical studies.

Here we approach this problem with the ``$\Lambda^*$-hypernuclei" picture proposed in Ref.~\cite{Arai:2007qj}, where the multi-baryon system with strangeness $S=-1$ is regarded as a composite of $\Lambda^*$ and nucleons, with $\Lambda^*$ being treated as an elementary particle.\footnote{In Ref.~\cite{Fink:1990hf}, the terminology ``$\Lambda^{*}$-hypernuclei'' was introduced to refer to the $\bar{K}$-nuclei, although explicit calculation in this picture was not performed.}
In the variational studies\cite{Yamazaki:2007cs,Dote:2008in,Dote:2008hw}, it is found that a $\bar{K}N$ pair in the $\bar{K}NN$ bound state has large overlap with $\Lambda^{*}$ in vacuum, and thus looks like a $\Lambda^* N$ bound system.
Therefore, the $\Lambda^*$-hypernuclei might provide an alternative description of the $\bar{K}$-nuclei.
The $\Lambda^*$-hypernuclei picture could make it easy to study the ground state of the few-body $\bar{K}$-nuclei in a similar way to the ordinary hypernuclei.
In addition, the explicit inclusion of the $YN$ channels is easier than the $\bar{K}NN$-$\pi\Sigma N$ approach, since the number of particles is the same with the $\Lambda^*N$ system.

The basic theoretical input for the study of the $\Lambda^*$-hypernuclei is the interaction of $\Lambda^*$ and a nucleon.
However, for the lack of the information of $\Lambda^*$, the $\Lambda^*N$ interaction is not explicitly known.
Then, in the previous work for the $\Lambda^*N$ and $\Lambda^*NN$ system\cite{Arai:2007qj}, the $\Lambda^*N$ interaction is determined by a phenomenological one-boson-exchange potential to fit the results of FINUDA experiment.
On the other hand, with the help of the theoretical description of $\Lambda^*$, it is possible to construct the  $\Lambda^*N$ interaction and predict the properties of the $\Lambda^*$-hypernuclei.
For this purpose, the chiral unitary approach\cite{Kaiser:1995eg,Kaiser:1995cy,Oset:1998it,Oller:2000fj,Lutz:2001yb,Hyodo:2011ur} is a suitable model, since it successfully reproduces the $S=-1$ meson-baryon scattering observables together with the dynamically generated $\Lambda^*$ resonance, and gives the structure of $\Lambda^*$ explicitly.

Here we follow the strategy to search for the possible bound state of the $\Lambda^*$-hypernuclei by determining the $\Lambda^*N$ interaction with the chiral unitary approach.
In the present work, we focus on the $\Lambda^*N$ two-body system which is the most fundamental $\Lambda^*$-hypernuclei and reflects the property of a given $\Lambda^*N$ interaction pronouncedly.
To study the ground state of the $\Lambda^*N$ system, we construct the $\Lambda^*N$ one-boson-exchange potential and solve the Schr\"{o}dinger equation to obtain the bound state. 
The $\Lambda^*$ resonance is described by a superposition of two resonance pole states in the framework of the chiral unitary approach\cite{Jido:2003cb}.
It is known that such double-pole structure is a consequence of the attractive forces in both $\bar{K}N$ and $\pi\Sigma$ channels\cite{Hyodo:2007jq}.
Hence, we have the two-component $\Lambda^*N$ system with channel coupling among the components.

We proceed with the paper as follows.
In Sec.~\ref{sec:model}, we construct the $\Lambda^* N$ potential by extending the J\"{u}lich $YN$ potential\cite{Holzenkamp:1989tq,Reuber:1992dh,Reuber:1993ip} with the properties of $\Lambda^*$ being constrained by the chiral unitary approach.
We show how the feature of the microscopic structure of $\Lambda^*$ is converted into the potential model.
The obtained $\Lambda^*N$ potential and numerical results of the bound state of the $\Lambda^*N$ system are shown in Sec.~\ref{sec:results}.
In Sec.~\ref{sec:discussion}, theoretical uncertainties within this model are discussed, and the conclusion is given in  Sec.~\ref{sec:conclusion}.


\section{Model}
\label{sec:model}

\subsection{$\Lambda^* N$ potential}
\label{sec:potential}

In the present work, the possible bound state of the $\Lambda^* N$ two-body system is searched for by constructing the potential in the coordinate space.
The $\Lambda^* N$ state is labeled by the total spin $S$ and the orbital angular momentum $L$ as $|S,L\rangle$.
Since the spin of $\Lambda^*$ is $1/2$, the total spin of the $\Lambda^* N$ system can be $S=0$ or $S=1$.
We only consider the $L=0$ component as a candidate of the ground state.
In this case, the tensor and the spin-orbit terms in the $\Lambda^* N$ potential do not contribute and thus we are left with the central force with spin-spin terms.
In the $NN$ scattering with spin $S=1$, the mixing of the $d$-wave state due to the tensor force plays an important role to develop the bound state, deuteron, while in the present case, the $d$-wave mixing may not contribute so strongly because the pion exchange is absent in the leading order $\Lambda^* N$ interaction.

  \begin{figure}[tb]
  \begin{center}
  \subfigure[Isoscalar exchange]{
  \includegraphics*[scale=0.6]{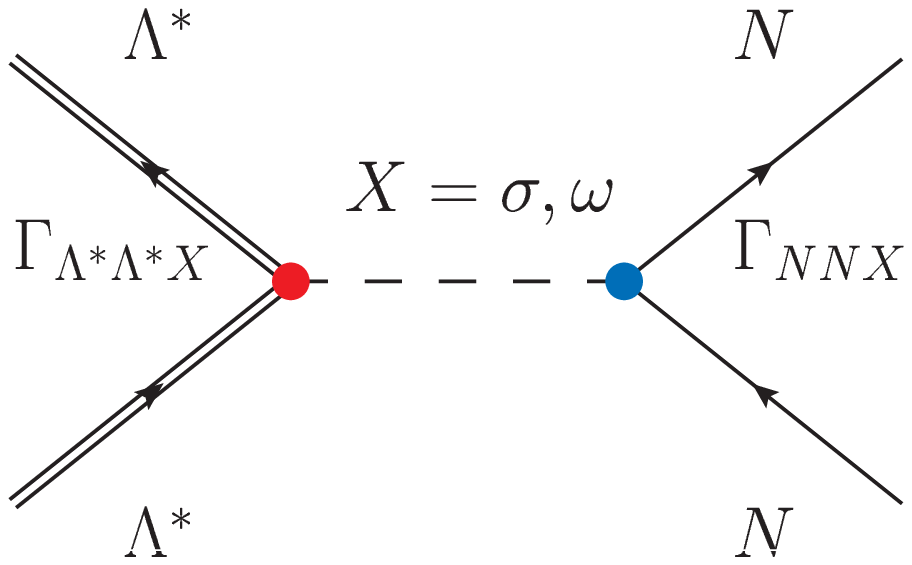}
  \label{fig:OBEP_X}}
  \subfigure[$\bar{K}$ exchange]{
  \includegraphics*[scale=0.6]{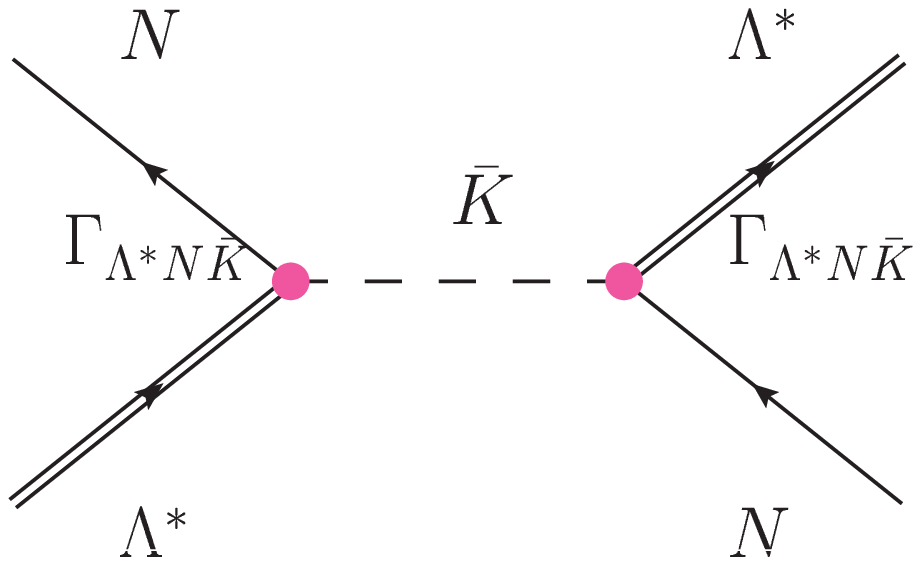}
  \label{fig:OBEP_K}}
  \end{center}
  \caption{The $\Lambda^*N$ potential in the one-boson-exchange picture. 
           The exchanged mesons are isoscalar mesons $X(X=\sigma, \omega)$ and $\bar{K}$. 
           The $\Gamma_{\Lambda^* N \bar{K}}$ and the $\Gamma_{NN X}$ vertices are determined by the chiral unitary approach and the J\"{u}lich potential respectively, while the remaining vertices $\Gamma_{\Lambda^* \Lambda^* X}$ are determined in Sec.~\ref{sec:estimation}.
           }
  \label{fig:OBEP}
  \end{figure}

In order to construct the $\Lambda^* N$ potential, we adopt the microscopic structure of $\Lambda^*$ given by the chiral unitary approaches of Refs.~\cite{Hyodo:2002pk,Hyodo:2003qa} (HNJH model).
There, the $\Lambda^*$ resonance is generated dynamically through the channel coupling of $\bar{K}N$, $\pi\Sigma$, $\eta \Lambda$ and $K\Xi$ scatterings.
The interaction vertices are given by the Weinberg-Tomozawa term, which is the leading order piece of chiral perturbation theory. 
The flavor structure of the Weinberg-Tomozawa interaction is identical with the heavy mass limit of the vector meson exchange with flavor SU(3) symmetric couplings.
The predicted scattering amplitude contains two resonance poles in the region of $\pi \Sigma$ and $\bar{K}N$ thresholds, both of which contribute to the resonance-like behavior identified as the $\Lambda^*(1405)$ resonance.
In our model, we describe the higher (lower) energy state of two poles of the $\Lambda^*$ resonance, as $\Lambda^*_1$ ($\Lambda^*_2$) which appears at 1427 MeV (1400 MeV) in the HNJH model where we interpret the real part of the pole position as the mass of $\Lambda^*_a$.
Accordingly, the $\Lambda^* N$ system also consists of two components, $\Lambda^*_1 N$ and $\Lambda^*_2 N$, and we solve the two-channel coupled Schr\"{o}dinger equation given by
   \begin{eqnarray}
   H\psi_{\Lambda^*N}&=&E\psi_{\Lambda^*N} ,
   \label{eq:Schrodinger}
   \end{eqnarray}
with the wave function
   \begin{eqnarray}
   \psi_{\Lambda^*N}&=&\left(
                     \begin{array}{c}
                     \psi_1 \\
                     \psi_2 \\
                     \end{array}
                     \right), 
   \label{eq:wave_function}
   \end{eqnarray}
where each component $\psi_a(a=1,2)$ corresponds to the wave function of each $\Lambda^*_a N$ state.
Hamiltonian $H$ is written as a summation of the kinetic energy $T$ and potential $V$, which are $2 \times 2$ matrices, given by
   \begin{eqnarray}
   H=T+V,
   \end{eqnarray}
with
   \begin{eqnarray}
   T&=&\left(
     \begin{array}{cc}
     T_1 +\Delta M & 0 \\
     0 & T_2 \\
     \end{array}
     \right), 
     \label{eq:kinetic_matrix} \\
   V&=&\left(
     \begin{array}{cc}
     V_{11} & V_{12} \\
     V_{21} & V_{22} \\
     \end{array}
     \right), 
     \label{eq:potential_matrix}
   \end{eqnarray}
where $\Delta M=M_{\Lambda^{*}_{1}}-M_{\Lambda^{*}_{2}}$ is the mass difference between channel 1 and channel 2.
$T_a$ is the kinetic energy of the $\Lambda^*_a$ state, given by
   \begin{eqnarray}
   T_a=-\frac{1}{2\mu_a}\vec{\nabla}^2
   \ ,
   \end{eqnarray}
with the reduced mass $\mu_a=M_N M_{\Lambda^*_a}/(M_N + M_{\Lambda^*_a})$, where $M_N$ and $M_{\Lambda^*_a}$ stand for the masses of the nucleon and $\Lambda^*_a$.
Each diagonal component of the potential matrix, $V_{aa}(a=1,2)$, is the potential of the $\Lambda^*_a N$ state, while off-diagonal components, $V_{12}$ and $V_{21}$, lead the transition between the $\Lambda^*_1 N$ and $\Lambda^*_2 N$ state.

To construct the $\Lambda^{*}N$ potential, we employ the J\"{u}lich potential (Model A) which is a typical one-boson-exchange potential including the hyperons\cite{Holzenkamp:1989tq,Reuber:1992dh,Reuber:1993ip}.
In the meson exchange diagrams in Fig.~\ref{fig:OBEP_X}, the exchanged mesons should be isoscalar, since the isospin of $\Lambda^*$ is zero.
The scalar $\sigma$ and the vector $\omega$ exchanges are taken into account, while the pseudoscalar $\eta$ has been omitted as its coupling to the nucleon is small.
We further consider the exchange potential $\Lambda^*N\to N\Lambda^*$ due to $\bar{K}$ exchange given by the diagram in Fig.~\ref{fig:OBEP_K}. 
So the $\Lambda^* N$ potential $V(r)$ can be written as the sum of three contributions
   \begin{eqnarray}
   V(r)=
   V_\sigma (m_{\sigma},r) + V_\omega (m_{\omega},r) + V_{\bar{K}} (m_{\bar{K}},r)
   \label{eq:lambda_potential}
   \ .
   \end{eqnarray}
The explicit forms of the $\Lambda^* N$ potential are given in \ref{sec:obep}.
      
   \begin{table}[tb]
   \caption{Coupling strengths in isospin basis and subtraction constants of $\Lambda^*_1(1427)$ and $\Lambda^*_2(1400)$ in the HNJH model\cite{Hyodo:2002pk,Hyodo:2003qa}.}
   \begin{center}
   \begin{tabular}{c|rrc}
   \hline \hline
                    & $g^{(i)}_{\Lambda^*_1 BM} $ & $g^{(i)}_{\Lambda^*_2 BM} $ & $a \left( \mu=630{\rm MeV} \right)$ \\ \hline
   $\pi\Sigma (i=1)$      &  $-0.69-1.41i$   & $ 2.44-1.75i$  &   $-1.96$  \\
   $\bar{K} N (i=2)$      &  $ 2.63+0.89i$   & $-1.03+1.93i$  &   $-1.96$  \\ \hline
   \end{tabular}
   \label{table:parameter_HNJH}
   \end{center}
   \end{table}
   
As shown in Fig.~\ref{fig:OBEP}, the coupling constants in the $\Lambda^*N$ potential are classified into three types; the $NNX$$(X=\sigma,\omega)$ vertices ($\Gamma_{NNX}$), the $\Lambda^*N\bar{K}$ vertex ($\Gamma_{\Lambda^* N\bar{K}}$), and the $\Lambda^*\Lambda^*X$ ($\Gamma_{\Lambda^* \Lambda^* X}$) vertices.
For the $\sigma$ and the $\omega$ exchanges, the $\Gamma_{NNX}$ vertices are determined by the $NN \sigma$ and $NN \omega$ couplings in the J\"{u}lich model.
The $\Gamma_{\Lambda^* \Lambda^* X}$ vertices include the unknown $\Lambda^*_a \Lambda^*_b \sigma$ and $\Lambda^*_a \Lambda^*_b \omega$ couplings and then they are estimated in section \ref{sec:estimation} based on chiral dynamics.
The $\Lambda^*$ couplings to the meson-baryon channel $g^{(i)}_{\Lambda^*_a BM}$ ($i=1$ for $\pi\Sigma$ and $i=2$ for $\bar{K}N$) can be extracted from the residues of the poles in the chiral unitary model whose numerical values are listed in Table~\ref{table:parameter_HNJH}.
The coupling constants are obtained as complex values because of the resonance nature of $\Lambda(1405)$.
To obtain the $\Lambda^* N \bar{K}$ coupling in the potential model, we have to convert it into the real number.
Then, considering the magnitude of the residues of the poles reflects the coupling strength, we shall identify the absolute value of $g^{(2)}_{\Lambda^*_a BM}$ as the coupling constant.
In addition, the coupling constants are obtained in isospin basis in Refs.~\cite{Hyodo:2002pk,Hyodo:2003qa}, while the coupling constants in the J\"{u}lich model are given in the particle basis as
   \begin{eqnarray}
   {\cal L}^{int} &=& g_{\Lambda^*_{a} N \bar{K}} \bar{\Lambda^*_{a}} pK^{-} +
   g_{\Lambda^*_{a} N \bar{K}} \bar{\Lambda^*_{a}} n\bar{K}^{0} +h.c.,
   \end{eqnarray}
where $h.c.$ denotes the hermite conjugate.
In order to be consistent with the normalization in the J\"{u}lich model, the $\Lambda^*_{a} N \bar{K}$ coupling constant should be translated into the particle basis.
Since $\Lambda^*$ has isospin $I=0$, the following relation
   \begin{eqnarray}
   \left[
   \bar{K} N
   \right]_{I=0}
   =
   \frac{K^- p +  \bar{K}^0 n}{\sqrt{2}},
   \end{eqnarray}
leads to a requisite factor $1/\sqrt{2}$ for translation of basis.
Therefore, we use the $\Lambda^*_{a} N \bar{K}$ coupling constant $g_{\Lambda^*_{a} N \bar{K}}$ in the $\Lambda^{*}N$ potential as
   \begin{eqnarray}
   g_{\Lambda^*_{a} N \bar{K}}
   =
   \frac{\left| g^{(2)}_{\Lambda^*_a BM} \right| }{\sqrt{2}}.
   \label{eq:coupling}
   \end{eqnarray}
Note that the $\bar{K}$ exchange contribution is completely determined for a given $\Lambda^* N \bar{K}$ coupling.
   
\subsection{$\bar{K}$ exchange contribution}
\label{sec:K}

Let us take a close look at the $\bar{K}$-exchange term.
Because $\bar{K}$ is a pseudoscalar meson and the parity of $\Lambda^* $ is odd, the $\Lambda^* N \bar{K}$ coupling is a scalar type.
So the $\bar{K}$ exchange contribution has almost the same form as the scalar meson exchange contribution, but it should be multiplied by the following spin exchange factor
   \begin{eqnarray}
   -\frac{1+(\vec{\sigma}_1\cdot\vec{\sigma}_2)}{2}
   \to
   \left\{
   \begin{array}{cc}
   1&(S=0)\\
   -1&(S=1)\\
   \end{array}
   \right.
   \label{eq:spin_exchange}
   .
   \end{eqnarray}
Due to this factor and the attractive nature of the scalar exchange, the $\bar{K}$ exchange contribution is attractive for $^1S_0$ and is repulsive for $^3S_1$.
This spin-dependence is important for determining the spin of the ground state of the $\Lambda^* N$ bound system.

Because of the large mass difference between $\Lambda^*$ and $N$ ($ M_{\Lambda^*} - M_N \sim 465$ MeV), we should not ignore the energy transfer $k^0$ in contrast to the ordinary $YN$ interaction.
The effect of non-zero energy transfer is approximately taken into account by an effective $\bar{K}$ mass, assuming that the baryons are static.
Following Ref.~\cite{Arai:2007qj}, in the $\bar{K}$ propagator, we use the effective mass given by
   \begin{eqnarray}
   \tilde{m}_{\bar{K}} = \sqrt{m_{K}^2-k_0^2} \simeq \sqrt{m_{K}^2-(M_{\Lambda^*}-M_N)^2}\ , 
   \label{eq:effective_mass}
   \end{eqnarray}
instead of the physical $\bar{K}$ mass, $m_{\bar{K}}=495$~MeV.
Note that $\tilde{m}_{\bar{K}}$ depends on the mass of $\Lambda^*$, so we have different effective masses in each component of the $\Lambda^*N$ potential.
For the off-diagonal component of the potential which leads to the mixing of $\Lambda^*_1N$ and $\Lambda^*_2N$, we use the average mass $\bar{M}_{\Lambda^*}=\left( M_{\Lambda^*_1}+M_{\Lambda^*_2} \right)/2$.
Specifically, for the diagonal component $V_{11}$($V_{22}$), $\tilde{m}_{\bar{K}} = $91~MeV (184~MeV), while for the off-diagonal component, $\tilde{m}_{\bar{K}} = $146~MeV.
In general, when $M_{\Lambda^*}$ approaches the $\bar{K}N$ threshold at $1435$~MeV, the effective mass $\tilde{m}_{\bar{K}}$ becomes small and the $\bar{K}$ exchange contribution is enhanced.

In the present work, we determine the $\Lambda^*N\bar{K}$ coupling constant by the residue of the scattering amplitude in the chiral unitary model.
The HNJH model\cite{Hyodo:2002pk,Hyodo:2003qa} leads to the coupling strength $g^2_{\Lambda^*NK}/4\pi\sim 0.2$-0.3.
This is almost an order of magnitude larger than the value in Ref.~\cite{Arai:2007qj}, $g^2_{\Lambda^*NK}/4\pi= 0.064$ which is determined by the decay width of the $\Lambda^*\to \pi\Sigma$ process and the SU(3) relation, with the assumption that $\Lambda^*$ belongs to the flavor singlet.
The difference can be understood by the structure of $\Lambda^*$; in the chiral unitary model, the main component of $\Lambda^*$ is the $\bar{K}N$ bound state and hence it has a strong coupling to the $\bar{K}N$ state~\cite{Hyodo:2007jq}.
From the group theoretical point of view, this is a consequence of the strong SU(3) violation in $\Lambda^*$, due to the variation of the threshold energies.
In any event, the stronger $\Lambda^*N\bar{K}$ coupling than the previous work will enhance the $\bar{K}$ exchange contribution in the $\Lambda^*N$ potential.

\subsection{Estimation of the $\Lambda^* \Lambda^* \sigma$ and $\Lambda^* \Lambda^* \omega$ couplings}
\label{sec:estimation}

  \begin{figure}[tb]
  \begin{center}
  \subfigure[two baryons coupling]{
  \includegraphics*[scale=0.6]{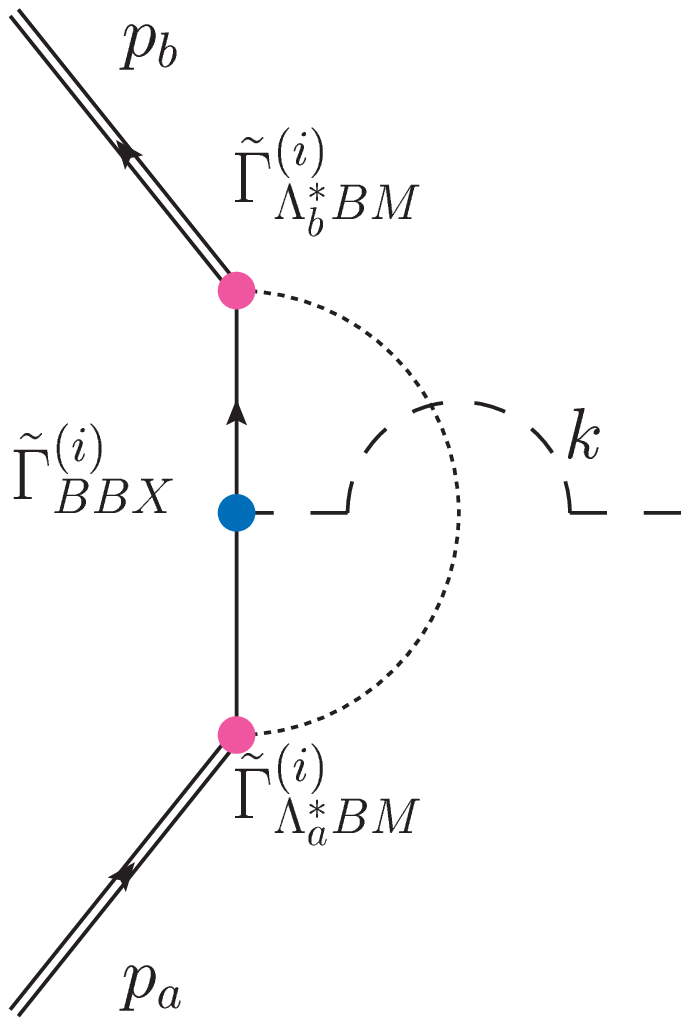}
  \label{fig:est_B}}
  \subfigure[two mesons coupling]{
  \includegraphics*[scale=0.6]{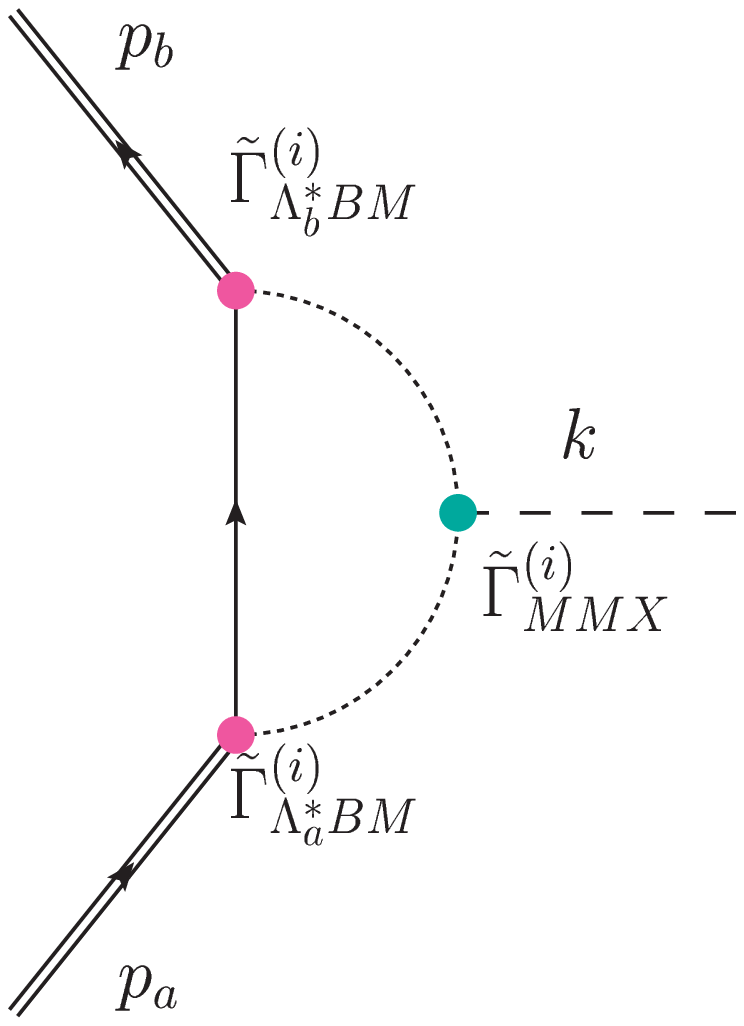}
  \label{fig:est_M}}
  \end{center}
  \caption{Microscopic description of the $\Lambda^* \Lambda^* X$ vertices ($X = \sigma, \omega$). 
           $\Lambda^*$ and the exchanged meson $X$ are represented by the double lines and the dashed lines. 
           The dotted lines and the solid lines denote the intermediate meson and baryon ($\pi\Sigma$ or $\bar{K}N$), respectively.
           }  
  \label{fig:estimation}
  \end{figure}

The key to construct the $\Lambda^*N$ potential in terms of meson-exchange diagrams is to evaluate the $\Lambda^* \Lambda^* X$ ($X=\sigma,\omega$) coupling constant.
Although it is difficult to directly extract the $\Lambda^* \Lambda^* X$ coupling from the experimental data, we can estimate the strength with help from the microscopic structure of $\Lambda^*$ obtained by the chiral unitary approach.
Here we treat two $\Lambda^*$ poles generated in the coupled-channel multiple scattering amplitude in the strangeness $S=-1$ and isospin $I=0$ channel.
There are four meson-baryon channels ($\pi\Sigma, \bar{K}N, \eta\Lambda, K\Xi$), but we deal only with the $\pi \Sigma$ and the $\bar{K} N$ components since they are the major components in the $\Lambda^*$ resonance, and the $\eta \Lambda$ and $K \Xi$ contributions will be suppressed in the estimation of the $\Lambda^* \Lambda^* X$ couplings.
It is shown that $\Lambda^*$ is dominated by the meson-baryon component\cite{Hyodo:2008xr}, so the exchanged meson $X$ couples to either the intermediate baryon or the intermediate meson in the multiple scattering as shown in Fig.~\ref{fig:estimation}.
The $\Lambda^*\Lambda^*X$ vertices, $\Gamma_{\Lambda^* \Lambda^* X}$, is given by the sum of these contributions:
   \begin{eqnarray}
   \Gamma_{\Lambda^* \Lambda^* X} = \Gamma_{\Lambda^* \Lambda^* X}^B + \Gamma_{\Lambda^* \Lambda^* X}^M
   \label{eq:estimation}
   \ ,
   \end{eqnarray}
where $\Gamma_{\Lambda^* \Lambda^* X}^B$ and $\Gamma_{\Lambda^* \Lambda^* X}^M$ stand for contributions of the diagrams in Fig.~\ref{fig:est_B} and ~\ref{fig:est_M} respectively. 
The coupling constants are obtained by taking the soft limit $\vec{k}\to\vec{0}$ in these diagrams.
Detailed calculations of the loop diagrams are shown in \ref{sec:coupling}.

The vertices which appear in the diagrams in Fig.~\ref{fig:estimation} are classified into three types.
First, the vertices $\tilde{\Gamma}^{(i)}_{\Lambda^* BM}$ in both diagrams are given by the chiral unitary approach.
The second type is the meson-baryon couplings, given according to the J\"{u}lich model, $\tilde{\Gamma}^{(i)}_{BBX}$ where $X$ denotes the type of the meson and $(i)$ designates two $\Lambda^*$ components, $\bar{K} N$ and $\pi \Sigma$.
The couplings of $\sigma$ and $\omega$ to the intermediate mesons are the third type vertices in vertices, $\tilde{\Gamma}_{MMX}^{(i)}$.
We follow Refs.~\cite{Hyodo:2002pk,Hyodo:2003qa} for $\tilde{\Gamma}^{(i)}_{\Lambda^* B M}$ and $\tilde{\Gamma}^{(i)}_{BBX}$ are taken from the J\"{u}lich(A) potential with the given form factor.
The $\tilde{\Gamma}^{(i)}_{MMX}$ vertices are determined by the property of the $\sigma$ meson in Refs.~\cite{Kaminski:2009qg,Ishida:1995xx,Oller:1998zr}, as discussed in \ref{sec:coupling}.

In the present model, we construct the $\Lambda^*N$ potential by treating the $\Lambda^*$ resonance as the fundamental degrees of freedom, so the coupling constants should be real values for the hermite interaction Lagrangian.
However, the couplings estimated by the loop diagrams in Fig.~\ref{fig:estimation} are in general complex values, because of the complex vertices $\tilde{\Gamma}_{\Lambda^* BM}^{(i)}$ and the contributions from the lower energy $\pi\Sigma$ loop. 
Since the imaginary part from the $\pi\Sigma$ loop represents the decay process of $\Lambda^{*}$ through the meson coupling, we avoid this contribution by taking only the principal value of the loop integral into account. 
On the other hand, we use the complex couplings for the $\tilde{\Gamma}_{\Lambda^* BM}^{(i)}$ vertices to correctly incorporate the relative phase between $\pi\Sigma$ and $\bar{K}N$ channels. 
After coherent summation of the $\pi\Sigma$ and $\bar{K}N$ channels, we take the absolute value of the amplitude to derive the real-valued coupling constant in the potential model in the same manner as the $\Lambda^* N \bar{K}$ case.
We have taken the sign of $\sigma \Lambda^* \Lambda^*$ coupling to be the same as that of $\sigma NN$ vertex.
It is natural because the scalar meson coupled to the internal structure of the baryons brings no extra phase to the $\sigma$ - baryon couplings.

\subsection{Form factor}
\label{sec:Form factor}

For a hadron having the finite size, coupling strengths between the hadron and the exchanged meson depend on the relative distance of the system.
In the momentum space, the coupling constant $g$ is described as a function of the momentum transfer $\vec{k}$.
This effect is included as a monopole type form factor $F(\vec{k})$ at each vertex, following the J\"{u}lich model\cite{Holzenkamp:1989tq}
   \begin{eqnarray}
   g \to gF(\vec{k})
      \ , \\
   F(\vec{k})=\frac{\Lambda^2-m^2}{\Lambda^2+\vec{k}^2}
   \ ,
   \label{eq:form_factor}
   \end{eqnarray}
where $\Lambda$ is the cut-off parameter and $m$ is the mass of the exchanged meson.

We use the same cut-off parameter as the J\"{u}lich potential for the $NNX$ vertices.
The $\Lambda^* \Lambda^* X$ vertices reflect the size of $\Lambda^*$ which is considered to be larger than the nucleon, due to the hadronic molecule structure\cite{Sekihara:2008qk,Sekihara:2010uz}.
We take into account the difference of $\Lambda^*$ and a nucleon by a constant $c$ as
   \begin{eqnarray}
   c=\frac{\sqrt{\left<r^2\right>_{\Lambda^*}}}{\sqrt{\left<r^2\right>_N}}
   \label{eq:c}
   \ ,
   \end{eqnarray}
which leads to
   \begin{eqnarray}
   \Lambda_{\Lambda^*\Lambda^* X}=\frac{\Lambda_{NNX}}{c}
   \label{eq:cut_lamlamX}
   \ .
   \end{eqnarray}
Meanwhile, for the $\Lambda^* N \bar{K}$ case, we consider the cut-off of the $NN\pi$ vertex in the J\"{u}lich model as a benchmark of the cut-off of the pseudoscalar vertex.
Taking into account the fact that one of the external baryon is $\Lambda^*$, we use
   \begin{eqnarray}
   \Lambda_{\Lambda^* N \bar{K}}=\frac{\Lambda_{NN \pi}}{\sqrt{c}}
   \label{eq:cut_lamNK}
   \ .
   \end{eqnarray}
The charge radius of the nucleon is about 0.88~fm, while the mean-squared radius of $\bar{K}$ in $\Lambda^*$ is estimated to be $\sim$ 1.4~fm, when the decay channel is eliminated\cite{Sekihara:2010uz}.
Here we adopt $c=1.5$ for both $\Lambda^*_1$ and $\Lambda^*_2$ as a representative value for the numerical calculations.
We examine the $c$ dependence of the results in Sec.~\ref{sec:discussion}.

Introducing the form factors, the meson exchange contribution in Eq.~\ref{eq:lambda_potential} is replaced as
   \begin{eqnarray}
   V_\alpha (m_\alpha,r) \to V_\alpha (m_\alpha,r) - \frac{\Lambda_2^2-m_\alpha^2}{\Lambda_2^2-\Lambda_1^2}V_\alpha (\Lambda_1,r) + \frac{\Lambda_1^2-m_\alpha^2}{\Lambda_2^2-\Lambda_1^2}V_\alpha(\Lambda_2,r)
   \ ,
   \end{eqnarray}
where $\alpha$ denotes the isoscalar $X$ or $\bar{K}$, and the $\Lambda_{1,2}$ are the cut-off parameters for each vertex.
For the isoscalar exchange, $\Lambda_{1,2}$ are $\Lambda_{NNX}$ and $\Lambda_{\Lambda^* \Lambda^* X}$.
Whereas, in the $\bar{K}$ exchange where the same cut-off is applied to the two vertices, we adopt the prescription in the J\"ulich model by setting
   \begin{eqnarray}
   \Lambda_1=\Lambda_{\Lambda^* N \bar{K}}+\epsilon, \Lambda_2=\Lambda_{\Lambda^* N \bar{K}}-\epsilon
   \ ,
   \end{eqnarray}
with $\epsilon=10$ MeV such that $\epsilon / \Lambda_{\Lambda^* N \bar{K}} \ll 1$.


\section{Results}
\label{sec:results}

We first show the properties of the constructed $\Lambda^*N$ potentials in Sec.~\ref{sec:LamN_potential}, and then discuss the results of bound state solutions of the $\Lambda^* N$ system.
In order to study the effects of channel coupling, we first solve the $\Lambda^*_1 N$ and the $\Lambda^*_2 N$ systems separately in Sec.~\ref{sec:single}.
Next we turn on the mixing between the $\Lambda^*_1 N$ and the $\Lambda^*_2 N$ states in Sec.~\ref{sec:coupled}, and see how the two channels mix in the quasi-bound state.

\subsection{Properties of the $\Lambda^* N$ potential}
\label{sec:LamN_potential}

   \begin{table}[tb]
   \caption{Coupling constants and the effective $\bar{K}$ masses of the $\Lambda^{*}N$ potential with the HNJH model\cite{Hyodo:2002pk,Hyodo:2003qa}.}
   \begin{center}
   \begin{tabular}{c|rrrrr}
   \hline \hline
   & $g_{\Lambda^* N \bar{K}} / \sqrt{4\pi}$ & $g_{\Lambda^* \Lambda^* \sigma} / \sqrt{4\pi}$ & $g_{\Lambda^* \Lambda^* \omega} / \sqrt{4\pi}$ & $f_{\Lambda^* \Lambda^* \omega} / \sqrt{4\pi}$ & $\tilde{m}_{\bar{K}}$~(MeV) \\ \hline
   $\Lambda^*_1(1427)$ & $0.55$ & 5.04 & 18.13 & 11.93 &  91 \\
   $\Lambda^*_2(1400)$ & $0.44$ & 1.12 &  4.83 &  3.60 & 184 \\
   Transition          & -      & 2.01 &  8.22 &  5.53 & 146 \\
   \hline
   \end{tabular}
   \label{table:g_HNJH}
   \end{center}
   \end{table}

We show the numerical results of the estimated coupling constants of the $\Lambda^*\Lambda^* \sigma$ and the $\Lambda^* \Lambda^* \omega$ vertices in Table~\ref{table:g_HNJH}, which are used in the $\Lambda^*N$ potential.
The magnitude of each coupling constant for $\Lambda^*_1$ is larger than the corresponding coupling of $\Lambda^*_2$.
The difference between the couplings of $\Lambda^*_1$ and $\Lambda^*_2$ is attributed not only to the mass of $\Lambda^{*}_{a}$, but also to the coupling strengths to the $\pi \Sigma$ and the $\bar{K}N$ channels in the multiple scattering, as seen in Table~\ref{table:parameter_HNJH}.
The couplings to the $\omega$ meson are stronger than the coupling to the $\sigma$ for both the $\Lambda^*_1$ and $\Lambda^*_2$ cases, as well as the transition couplings.

  \begin{figure}[tb]
  \begin{center}
  \subfigure[$\Lambda^*_1 N$,$^1S_0$]{
  \includegraphics*[scale=0.7]{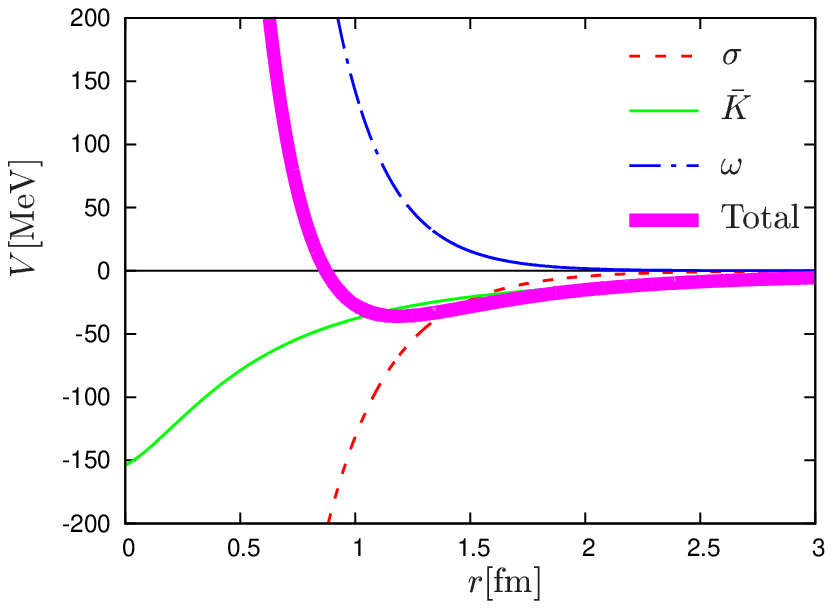}
  \label{fig:V_1_HNJH_S0}}
  \subfigure[$\Lambda^*_1 N$,$^3S_1$]{
  \includegraphics*[scale=0.7]{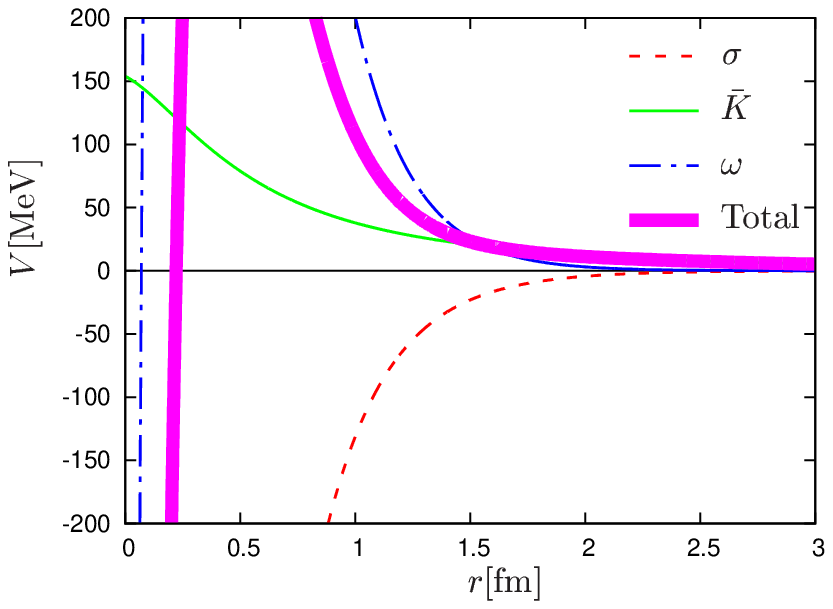}
  \label{fig:V_1_HNJH_S1}} \\
  \subfigure[$\Lambda^*_2 N$,$^1S_0$]{
  \includegraphics*[scale=0.7]{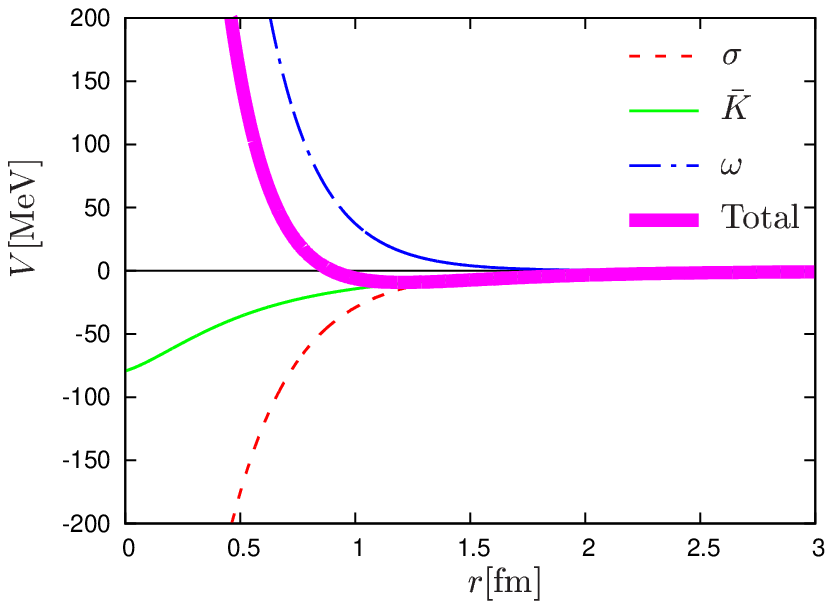}
  \label{fig:V_2_HNJH_S0}}
  \subfigure[$\Lambda^*_2 N$,$^3S_1$]{
  \includegraphics*[scale=0.7]{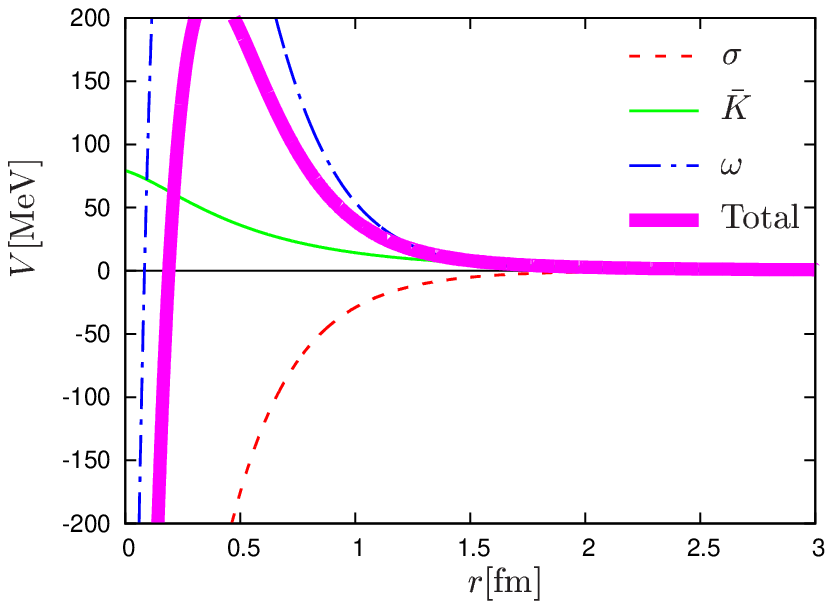}
  \label{fig:V_2_HNJH_S1}}
  \end{center}
  \caption{$\Lambda^*_a N$$(a=1,2)$ potentials for the $^1S_0$ and $^3S_1$ channels.
           Thick solid lines stand for the total potential and thin solid, dashed, and dash-dotted lines represent the $\bar{K}$, $\sigma$ and $\omega$ exchange contributions.}
  \label{fig:V_HNJH}
  \end{figure}

  \begin{figure}[tb]
  \begin{center}
  \subfigure[$\Lambda^*_1 N$,$^1S_0$]{
  \includegraphics*[scale=0.7]{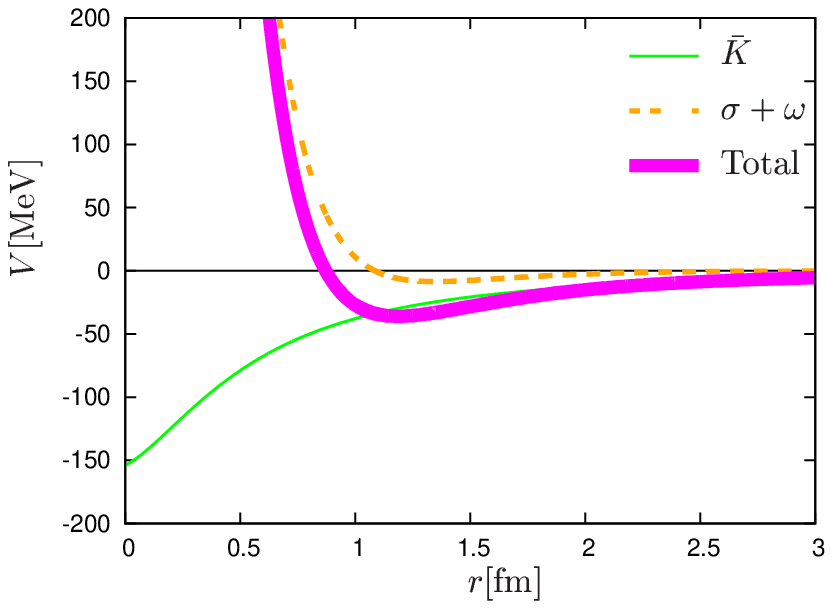}
  \label{fig:VV1_HNJH_S0}}
  \subfigure[$\Lambda^*_1 N$,$^3S_1$]{
  \includegraphics*[scale=0.7]{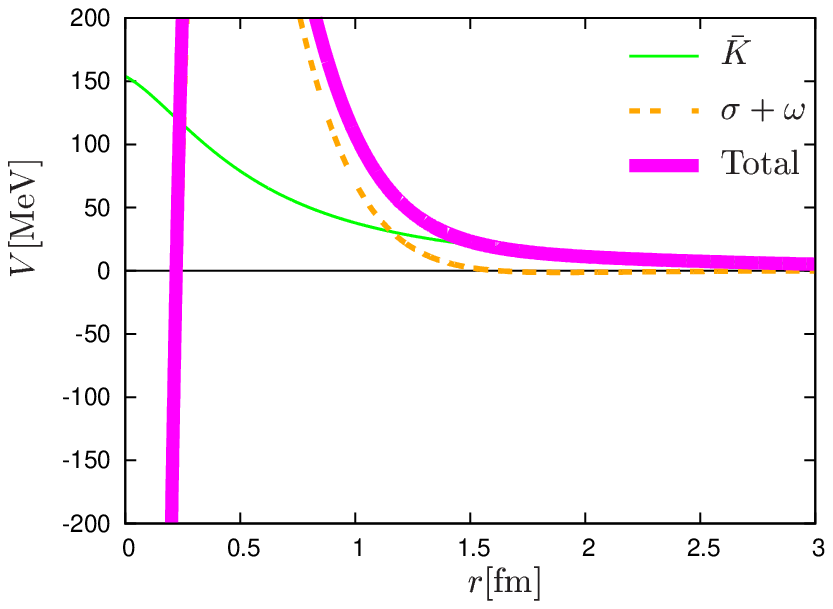}
  \label{fig:VV1_HNJH_S1}}
  \end{center}
  \caption{The $\Lambda^*_1 N$ potential for the $^1S_0$ and $^3S_1$ channels. 
           Thick solid lines stand for the total potential and thin solid and dashed lines represent the $\bar{K}$ and isoscalar exchange contributions.}
  \label{fig:VV_HNJH}
  \end{figure}

With these couplings, we plot in Fig.~\ref{fig:V_HNJH} the diagonal components of the $\Lambda^* N$ potential, $V_{11}$ and $V_{22}$ which represent the $\Lambda^*_1 N$ and $\Lambda^*_2 N$ interactions, for the $^1S_0$ and $^3S_1$ channels. We also show the individual contributions from $\sigma$, $\omega$ and $\bar{K}$ exchanges.
It can be seen that the potentials depend strongly on the total spin of the $\Lambda^*_a N$ system, while the qualitative feature of the $\Lambda^*_1 N$ potential is similar to the $\Lambda^*_2 N$ potential.
The contributions from the $\sigma$ and $\omega$ exchanges are stronger than that of the $\bar{K}$ exchange.
In the intermediate range region, however, there is a large cancellation between the attractive $\sigma$ exchange and the repulsive $\omega$ exchange as shown by the contribution from the isoscalar exchange which is the sum of the $\sigma$ and $\omega$ contributions in Fig.~\ref{fig:VV_HNJH}. 
As a consequence, the contribution from the $\bar{K}$ exchange is relatively important to determine the sign of the potential.
As noted in Sec. \ref{sec:K}, the $\bar{K}$ exchange is repulsive for the $S=1$ system, which results in the repulsive nature of the total potential except for the very short range region.
For the spin $S=0$ case, the $\bar{K}$ exchange contributes attractively and the $\Lambda^*N$ potential has an attractive pocket in the intermediate range and the repulsive core at short distance.

Before solving the Schr\"{o}dinger equation, let us study the bulk property of the $\Lambda^* N$ interaction by calculating the volume integral of the potential, which is the potential in momentum space at $p=0$
   \begin{eqnarray}
   V(p=0) = \int V(r)d^3r=4\pi\int_0^{\infty} r^2V(r)dr
   \ .
   \end{eqnarray}
The results of the volume integral are listed in Table~\ref{table:vol_int}.
   \begin{table}
   \caption{Volume integral of the $\Lambda^* N$ potential.}
   \begin{center}
   \begin{tabular}{c|rr}
   \hline \hline
                                                  & $S=0$  & $S=1$ \\ \hline
   $\int V_{11}(r)d^3r[{\rm MeV}\cdot{\rm fm}^3]$ & $-2602$ & $4560$  \\ 
   $\int V_{22}(r)d^3r[{\rm MeV}\cdot{\rm fm}^3]$ & $ -141$ & $ 899$  \\ 
   \hline
   \end{tabular}
   \label{table:vol_int}
   \end{center}
   \end{table}
For the spin $S=1$ case, it can be seen that both $\Lambda^*_1 N$ and $\Lambda^*_2 N$ potential are repulsive.
Since the integrand contains the $r^2$ factor, the short range attraction in $S=1$ does not affect the bulk property of the potential very much.
On the other hand, for the spin $S=0$ case, although each $\Lambda^*_a N$ potential is found to be attractive, the attraction of the $\Lambda^*_2 N$ potential is weaker than that of the $\Lambda^*_1 N$ potential.
This results indicate that the $\Lambda^*_2 N$ potential may not have the attraction enough to develop the bound state.
For all channels, the volume integral reflects the property of the long range part of the potential where the $\bar{K}$ exchange dominates, because of the light effective mass as discussed in Sec.~\ref{sec:K}.
In addition, the $\Lambda^*_1 N$ potential is stronger than the $\Lambda^*_2 N$ one, reflecting the stronger coupling constants and the longer range of the $\bar{K}$ exchange due to the lighter effective kaon mass.

\subsection{$\Lambda^*N$ bound state without channel mixing}
\label{sec:single}

\begin{figure}[tb]
  \begin{center}
  \subfigure[wave function $\psi(r)$]{
  \includegraphics*[scale=0.7]{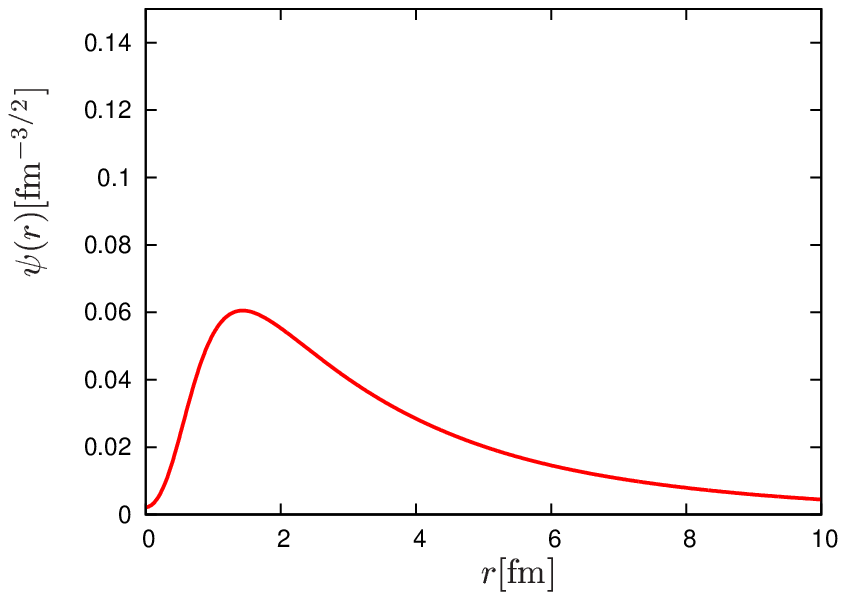}
  \label{fig:wf_HNJH}}
  \subfigure[density distribution $\left| r\psi(r) \right|^2$]{
  \includegraphics*[scale=0.7]{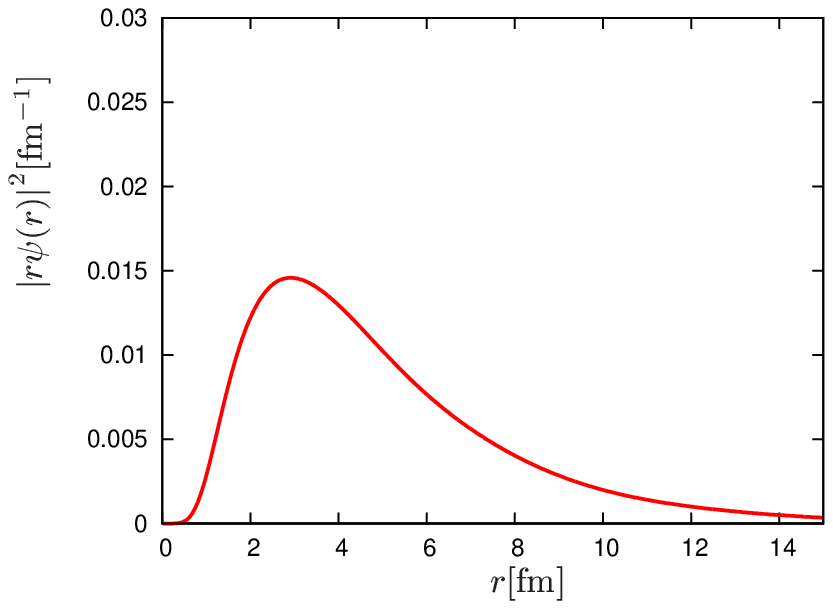}
  \label{fig:rwf_HNJH}}
  \end{center}
  \caption{The wave function of the $\Lambda^*_1 N$ system with the HNJH model in the coordinate space.}
  \label{fig:WF_HNJH}
  \end{figure}

First we search for bound state solutions of the $\Lambda^* N$ system in the $^1S_0$ and $^3S_1$ channels, solving the Schr\"{o}dinger equation with a variational approach called Gaussian Expansion Method (GEM)\cite{Hiyama:2003cu}.
In this section, we consider the $\Lambda^*_a N$ $(a=1,2)$ system in the single channel by switching off the off-diagonal component of the $\Lambda^* N$ potential, $V_{12}$ and $V_{21}$ in Eq.~\ref{eq:potential_matrix}.
For the $^3S_1$ case, we could not find any bound state solutions in either the $\Lambda^*_1 N$ or $\Lambda^*_2 N$ channels, in accordance  with the repulsive volume integral.
On the other hand, for the $^1S_0$ case, we find one bound state in the $\Lambda^*_1 N$ channel, while no bound state is found in the $\Lambda^*_2 N$ channel.
The mass of the bound state is
   \begin{eqnarray}
   M_{\Lambda^*_1N}
   =
   2365 {\rm ~MeV}
   \label{eq:energy_single}
   \ ,
   \end{eqnarray}
which corresponds to $\sim$ 1.0 MeV binding below the $\Lambda^*_1N$ threshold.
In the $\Lambda^*_aN$ potential model, the solution below the threshold is a stable bound state, although it can physically decay into the hadronic final states such as the three-body $\pi\Sigma N$ and $\pi\Lambda N$ channels as well as the two-body $\Sigma N$ and $\Lambda N$ states, if the corresponding phase space is available.
Note that the mass of the bound $\Lambda^*_1N$ system is higher than the threshold of the $\Lambda^*_2 N$ channel, 2339 MeV, so it will decay into $\Lambda^{*}_{2}N$ channel when the off-diagonal $\Lambda^{*}N$ potential is included. 
The wave function and the density distribution of the $\Lambda^*_1 N$ ground state with $^1S_0$ in the coordinate space are shown in Fig.~\ref{fig:WF_HNJH}.
It is found that the ground state of the $\Lambda^*_1 N$ system is a loosely bound state, peaking at the relative distance of $r \sim 3$ fm shown in Fig.~\ref{fig:rwf_HNJH}.
The wave function at the origin is suppressed by the short-range repulsion in the $\Lambda^{*}_{1}N$ potential shown in Fig.~\ref{fig:V_1_HNJH_S0}.
The mean distance is calculated as
   \begin{eqnarray}
   \sqrt{\langle r^2 \rangle}
   &=&
   \left[ \int d^3r\ r^2|\psi_{1}(r)|^2 \right]^{1/2} \nonumber \\
   &=&
   5.8~{\rm fm}.
   \end{eqnarray}
The size of the bound state is slightly larger than the deuteron, as is expected from the smaller binding energy.

\subsection{$\Lambda^*N$ bound state with channel mixing}
\label{sec:coupled}

Now we study the $\Lambda^* N$ system in the full channel coupling, and the lower energy $\Lambda^*_2 N$ threshold is chosen as the energy $E=0$. 
We find no bound state with $E<0$ in the coupled-channel Schr\"odinger equation, so the possible $\Lambda^{*}N$ system
will be a resonance state with $E>0$.
Due to the mixing with the $\Lambda^*_2 N$ continuum, the $\Lambda^*_1 N$ bound state obtained in the previous section will vary its energy.
In addition, if the $\Lambda^* N$ resonance exist above the $\Lambda^*_2 N$ threshold, the resonance has a finite decay width of the $\Lambda^*_1 N \to \Lambda^*_2 N$ decay process.
To see the mixing effect in accordance with off-diagonal components of the potential matrix, we introduce the parameter $\lambda(0 \leq \lambda \leq 1)$ which controls the mixing as
   \begin{eqnarray}
   V \to \left(
     \begin{array}{cc}
     V_{11} & 0      \\
     0      & V_{22} \\
     \end{array}
     \right)
     +\lambda
     \left(
     \begin{array}{cc}
     0      & V_{12}  \\
     V_{21} & 0       \\
     \end{array}
     \right),
   \label{eq:V_lambda}
   \end{eqnarray}
where the $\lambda=0$ case reproduces the single channel calculation performed in Sec.~\ref{sec:single}, while the $\lambda=1$ case corresponds to the full channel coupling.
Accordingly, we study the $\Lambda^* N$ system in the $^1S_0$ channel by varying the parameter $\lambda$.

To study the resonant $\Lambda^* N$ system, we use the real scaling method\cite{Hiyama:2005cf,RS} which is one of the techniques to search for a resonance in the continuum.
Before the calculation, we show how the real scaling works with the Gaussian Expansion Method (GEM).
In GEM, the wave function is expanded in a finite number of Gaussian basis and the explicit form of $\Lambda^{*}_{a}N$ wave function in the $s$-wave channel is given by
    \begin{eqnarray}
    \psi_a (r)
    &=&
    \frac{1}{\sqrt{4\pi}} \sum_{n=1}^{n_{max}} c^{(a)}_n N_n e^{-(r/r_n)^2},
    \end{eqnarray}
with
    \begin{eqnarray}
    N_n
    &=&
    \left\{ \frac{4}{\sqrt{\pi}} \left( \frac{2}{r_n^2} \right)^{3/2} \right\}^{1/2} , \\
    r_n
    &=&
    r_1 \left( \frac{r_{max}}{r_1} \right)^{\frac{n-1}{n_{max}-1}}.
    \end{eqnarray}
where $N_n$ is the normalization factor,  $r_n$ limits the spatial region of the wave function, $c^{(a)}$ is the coefficient of each basis for the $\Lambda^*_aN$ component and $n_{max}$ is the number of the basis functions.
We take $n_{max}= 20$, $r_{1}=0.05$fm and $r_{max}=10$fm.
Since 
the basis functions have the limited range, all the eigenvalues are discrete.
Now we introduce a scale parameter $\alpha$ as
    \begin{eqnarray}
    r_{max} \to \alpha r_{max},
    \end{eqnarray}
with $n_{max}$ and $r_1$ fixed.
Then, the energy eigenvalues will change accordingly.
Most eigenvalues will fall down towards the threshold, when $\alpha$ becomes larger.
This is called real scaling method.
If a sharp resonance exists at energy $E_R$, 
$E_R$ will be kept constant as the compact resonance state is not affected by the boundary.
It is, however, modified when one of the discrete (scattering) state crosses $E_R$. 
At the crossing, due to the mixing of the scattering state, the energy eigenvalue corresponding to the resonance $E_R$ is pushed away. 
The larger the mixing (coupling) to the scattering state is, the larger is the deviation.
One can estimate the decay width roughly from the deviation.

\begin{figure}[tb]
  \begin{center}
  \subfigure[$\lambda=0$]{
  \includegraphics*[scale=0.7]{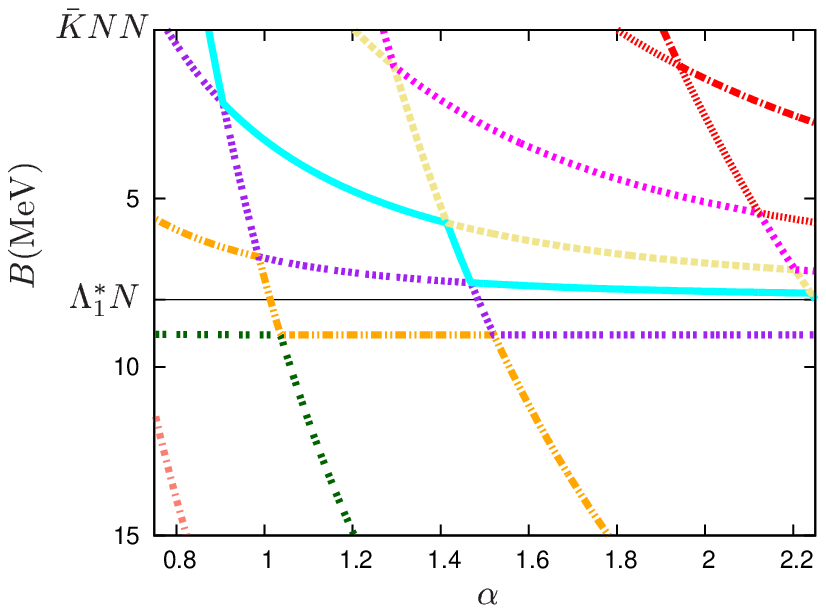}
  \label{fig:rs_lam0}}
  \subfigure[$\lambda=1/2$]{
  \includegraphics*[scale=0.7]{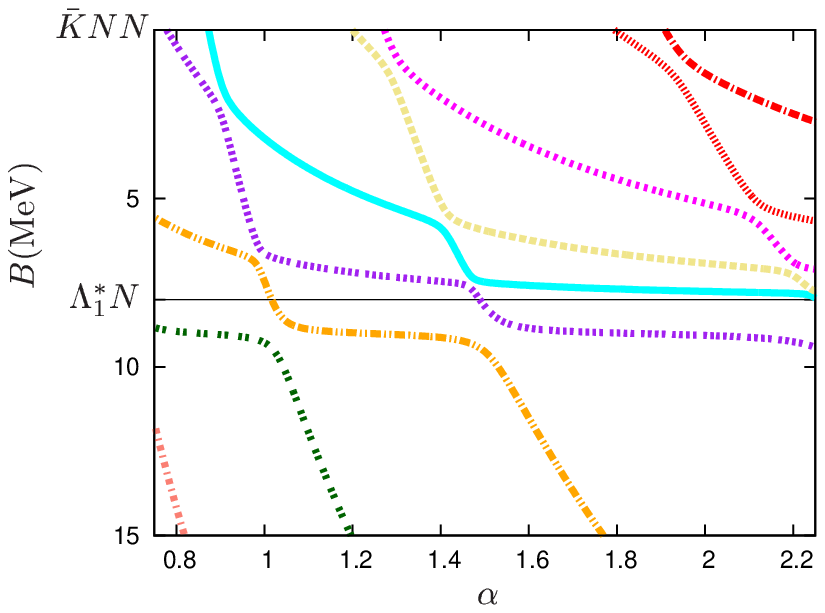}
  \label{fig:rs_lam05}}
  \subfigure[$\lambda=1$]{
  \includegraphics*[scale=0.7]{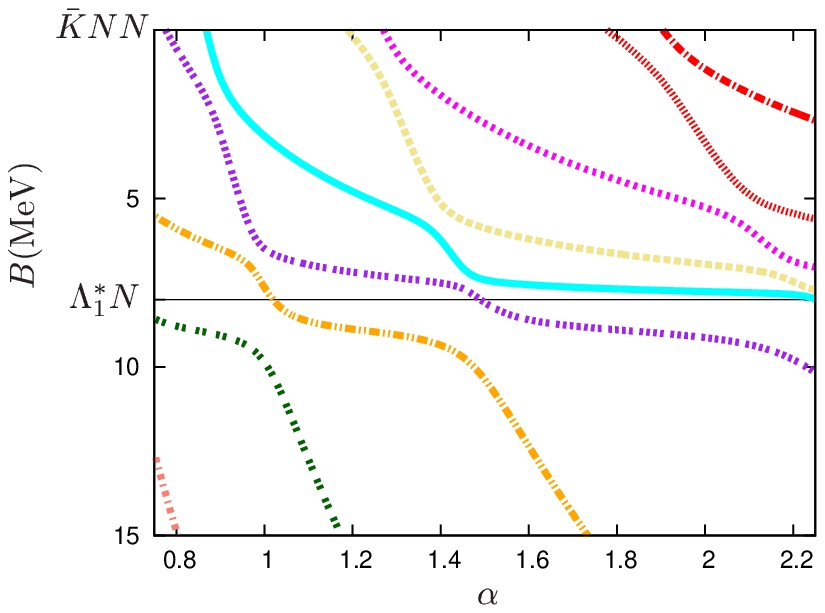}
  \label{fig:rs_lam1}}
  \end{center}
  \caption{Energy eigenvalues of the $\Lambda^* N$ system with the HNJH chiral unitary model,\cite{Hyodo:2002pk,Hyodo:2003qa} obtained by the real scaling method.
           The energy is given as the binding energy $B$ which is measured from the $\bar{K}NN$ threshold.
           With the $\bar{K}$, $N$ and $\Lambda^*_2$ mass, $m_{\bar{K}}$, $M_N$ and $M_{\Lambda^*_2}$, $B$ is defined as $B=(m_{\bar{K}}+2M_N)-(M_{\Lambda^*_2}+M_N)-E$.
           The parameter $\alpha$ controls the range of the wave function. }
  \label{fig:rs}
  \end{figure}

With the use of the real scaling method, we search for resonances in the full channel coupling. 
The results of the energy eigenvalues are shown as functions of the parameter $\alpha$ with  $\lambda=0,1/2,1$ in Fig.~\ref{fig:rs}.
In Fig.~\ref{fig:rs_lam0}, there are two classes of the $\alpha$-dependent scattering states: those which fall towards the $\Lambda^{*}_{1}N$ threshold and those which go to the lower energies.
The former (latter) levels correspond to the $\Lambda^{*}_{1}N$ ($\Lambda^{*}_{2}N$) scattering states. In addition, we observe one $\alpha$-independent eigenvalue below the $\Lambda^{*}_{1}N$ threshold, which represents the $\Lambda^*_1 N$ bound state obtained in the previous section.
By including half of the mixing effect as shown in Fig.~\ref{fig:rs_lam05}, the spectrum shows level repulsion at the crossing points and there is a plateau in between. 
The energy is close to the corresponding $\Lambda^{*}_{1}N$ bound state without channel mixing. This means that the bound state acquires the decay width through the channel coupling, while the energy of the resonance does not change very much.
For the full channel coupling case shown in Fig.~\ref{fig:rs_lam1}, we find one resonance as the $\Lambda^* N$ quasi-bound state with a decay width of the $\Lambda^*_1 N \to \Lambda^*_2 N$ process in the order of several MeV, in pretty much the same energy region as the $\lambda=0$ case.
In our present model, the $\Lambda^* N$ quasi-bound state is found to be strongly dominated by the $\Lambda^*_1 N$ bound state. This is in contrast with the $\Lambda^*$ resonance in the $\bar{K}N$-$\pi\Sigma$ system where the mixing of two components is important for the structure of the resonance.

\begin{figure}[tb]
  \begin{center}
  \subfigure[$\alpha=0.9$]{
  \includegraphics*[scale=0.7]{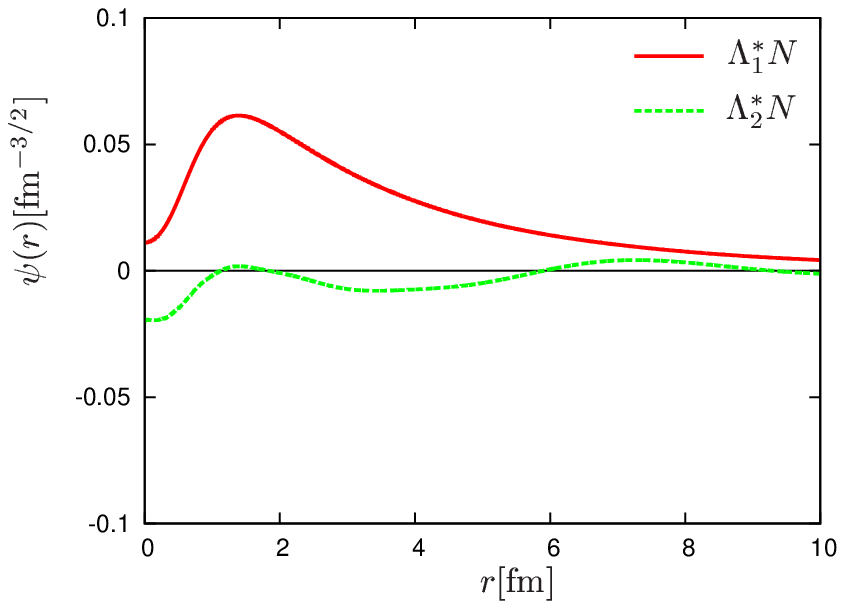}
  \label{fig:Cwf_09}}
  \subfigure[$\alpha=1.3$]{
  \includegraphics*[scale=0.7]{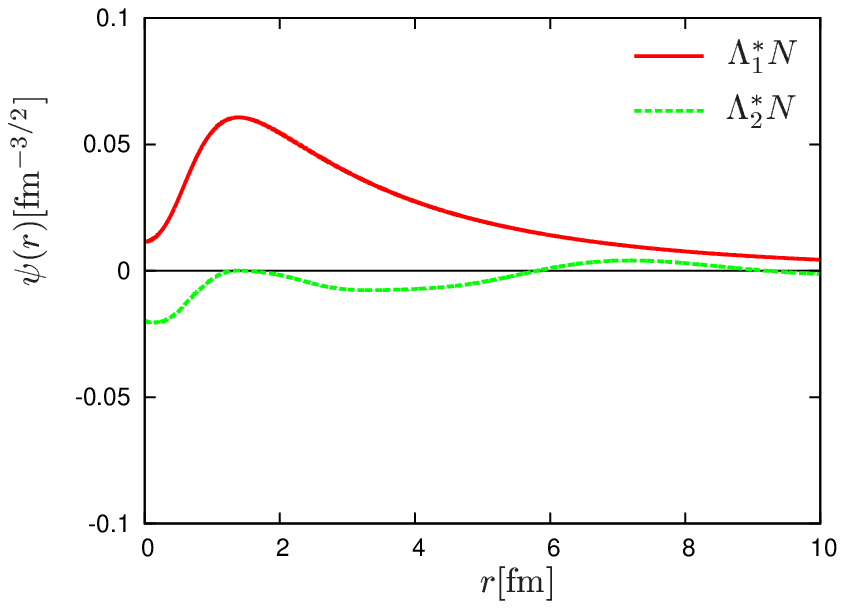}
  \label{fig:Cwf_13}} \\
  \subfigure[$\alpha=1.7$]{
  \includegraphics*[scale=0.7]{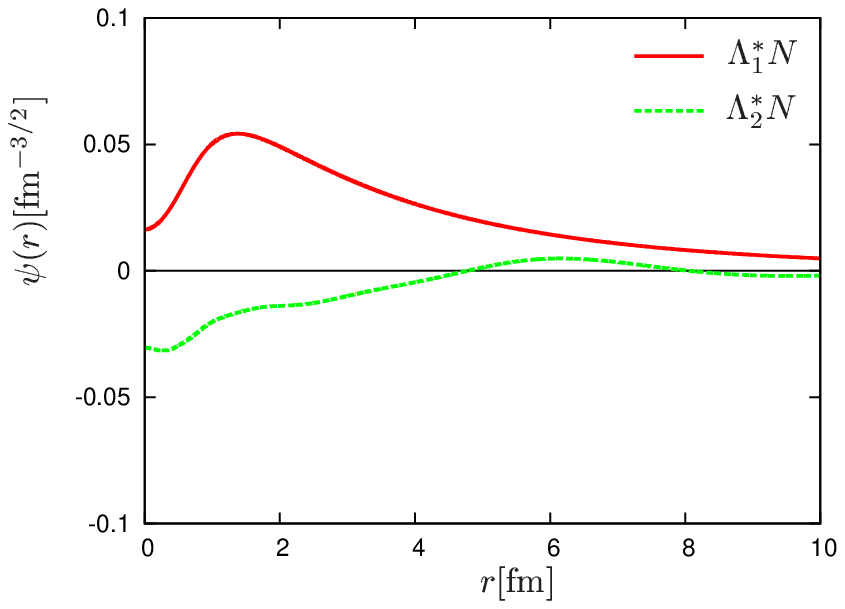}
  \label{fig:Cwf_17}}
  \subfigure[$\alpha=1.9$]{
  \includegraphics*[scale=0.7]{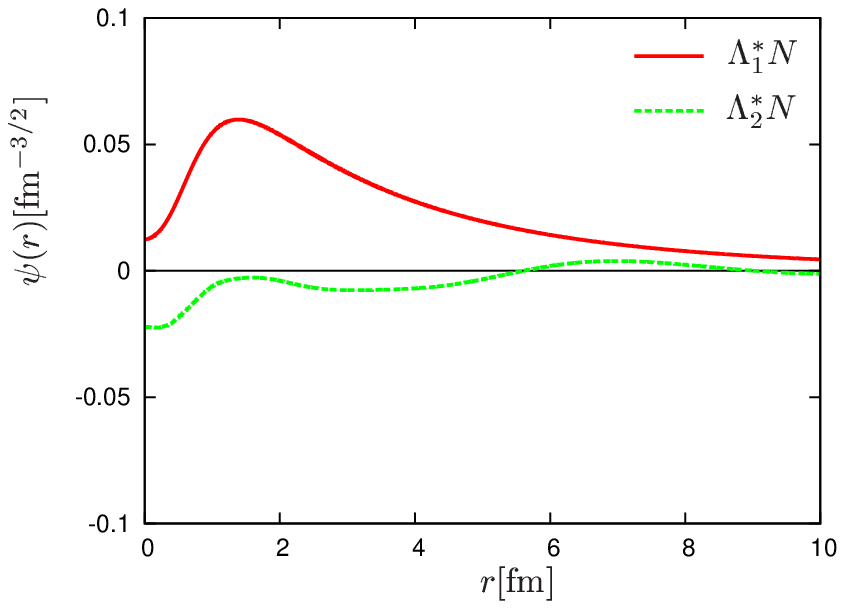}
  \label{fig:Cwf_19}}
  \end{center}
  \caption{Each component of the wave function in the $\Lambda^* N$ quasi-bound state.}
  \label{fig:Cwf}
  \end{figure}

In order to see the structure of the $\Lambda^* N$ quasi-bound state, we plot the wave functions of the resonant state for, $\alpha=0.9,1.3,1.7,1.9$ shown in Fig.~\ref{fig:Cwf}.
One sees that the $\Lambda^*_1 N$ component is found to be similar as the single channel case shown in Fig.~\ref{fig:WF_HNJH}, while the $\Lambda^*_2 N$ components do not contribute much.
We define the fraction $C_a$ of each component of the quasi-bound state, given by
  \begin{eqnarray}
   C_a
   =
   \frac{\int^R d^3r \left| \psi_a (r) \right|^2 }{\int^R d^3r \left| \psi_1(r) \right|^2 + \int^R d^3r \left| \psi_2(r) \right|^2},
   \end{eqnarray}
where $R$ corresponds to the range of the potential, and is set as 3 fm.
The results are given in Table~\ref{table:qb}.
The fractions $C_1$ show that the resonance state is dominated by $\Lambda^*_1 N$. 
The mean distance of the $\Lambda^*_1N$ component is given by
   \begin{eqnarray}
   \sqrt{\langle r^2 \rangle}
   =
   \left[
   \frac{\int d^3r r^2 \left| \psi_1(r) \right|^2 }{\int d^3r \left| \psi_1(r) \right|^2 }
   \right]^{1/2}.
   \label{eq:meandistance}
   \end{eqnarray}
The numerical values are listed in Table~\ref{table:qb}.
We find that the mean distance of the $\Lambda^* N$ quasi-bound state is about 5 fm, and is close to the result of the single channel case.
Among the four almost-arbitrary samplings of $\alpha$, the $\alpha=1.7$ case seems to have larger deviation from the single channel results.
The reason is not clear.

   \begin{table}[tb]
   \caption{The fraction of the $\Lambda^*_1 N$ component of the quasi-bound state $C_1$ and the mean distance $\sqrt{\langle r^2 \rangle}$ defined in Eq.~\ref{eq:meandistance} with several values of the parameter $\alpha$.}
   \begin{center}
   \begin{tabular}{c|cc}
   \hline \hline
   $\alpha$ & $C_1$ & $\sqrt{\langle r^2 \rangle}$~(fm) \\ \hline
   0.9      &  0.94 & 5.6 \\
   1.3      &  0.95 & 5.8 \\
   1.7      &  0.93 & 6.4 \\
   1.9      &  0.95 & 5.9 \\
   \hline
   \end{tabular}
   \label{table:qb}
   \end{center}
   \end{table}
 

\section{Discussion}
\label{sec:discussion}

So far we have studied the quasi-bound state in the $\Lambda^*N$ potential model. 
In the construction of the $\Lambda^*N$ potential, however, the experimental database is not sufficient to constrain all the details of the property of $\Lambda^*$\cite{Ikeda:2011dx}.
In addition, we do not have the exact value of the coupling between the exchanged meson to the pseudoscalar meson.
In order to estimate the theoretical uncertainties within the framework, here we discuss the possible ambiguities in the $\Lambda^*N$ potential and explore the model dependence.
It is also our aim to clarify the physical mechanism of the $\Lambda^*N$ binding and the limitation of the present approach, by carefully studying the response of the results to the potential parameters.

\subsection{Variants in the chiral unitary approach}
\label{sec:dis_chiral}

   \begin{table}[tb]
   \caption{
   Coupling strengths in isospin basis and subtraction constants of $\Lambda^*_1(1426)$ and $\Lambda^*_2(1390)$ in the ORB model\cite{Oset:2001cn}.
   }
   \begin{center}
   \begin{tabular}{c|rrc}
   \hline \hline
                    & $g^{(i)}_{\Lambda^*_1 BM} $ & $g^{(i)}_{\Lambda^*_2 BM} $ & $a \left( \mu=630{\rm MeV} \right)$ \\ \hline
   $\pi\Sigma(i=1)$      &    $ 0.42-1.4i$     & $-2.5-1.5i$  &   $-2.00$  \\
   $\bar{K} N(i=2)$      &    $ -2.5+0.94i$    & $ 1.2+1.7i$  &   $-1.84$  \\ \hline
   \end{tabular}
   \label{table:parameter_ORB}
   \end{center}
   \end{table}
   
   \begin{table}[tb]
   \caption{Coupling strengths in isospin basis and subtraction constants of $\Lambda^*_1(1434)$ and $\Lambda^*_2(1379)$ in the OM model\cite{Oller:2000fj}.}
   \begin{center}
   \begin{tabular}{c|rrc}
   \hline \hline
                    & $g^{(i)}_{\Lambda^*_1 BM} $ & $g^{(i)}_{\Lambda^*_2 BM} $ & $a \left( \mu=630{\rm MeV} \right)$ \\ \hline
   $\pi\Sigma(i=1)$      &  $-0.56-1.02i$   & $-1.76-0.62i$  &   $-2.23$  \\
   $\bar{K} N(i=2)$      &  $-1.74+0.63i$   & $ 0.86+0.70i$  &   $-2.23$  \\ \hline
   \end{tabular}
   \label{table:parameter_OM}
   \end{center}
   \end{table}
   
   \begin{table}[tb]
   \caption{Coupling strengths in isospin basis and subtraction constants of $\Lambda^*_1(1433)$ and $\Lambda^*_2(1388)$ in the BNW model\cite{Borasoy:2004kk,Borasoy:2005ie}.}
   \begin{center}
   \begin{tabular}{c|rrc}
   \hline \hline
                    & $g^{(i)}_{\Lambda^*_1 BM} $ & $g^{(i)}_{\Lambda^*_2 BM} $ & $a \left( \mu=630{\rm MeV} \right)$ \\ \hline
   $\pi\Sigma(i=1)$      &  $ 0.18-1.66i$   & $ 2.50-0.97i$  &   $-2.35$  \\
   $\bar{K} N(i=2)$      &  $ 2.25+1.32i$   & $-1.68+1.52i$  &   $-1.86$  \\ \hline
   \end{tabular}
   \label{table:parameter_BNW}
   \end{center}
   \end{table}

We first examine several variants in the chiral unitary approach to determine the properties of $\Lambda^*$.
In Refs.~\cite{Oller:2000fj,Oset:2001cn,Borasoy:2004kk,Borasoy:2005ie}, the meson-baryon scattering in the strangeness $S=-1$ sector and the $\Lambda^*$ resonance are studied. 
We call these models OM\cite{Oller:2000fj}, ORB\cite{Oset:2001cn} and BNW\cite{Borasoy:2004kk,Borasoy:2005ie}.
Constrained by the experimental data of the $K^-p$ scattering, all models found two $\Lambda^*$ poles in the scattering amplitude, while the pole positions and the coupling strengths of $\Lambda^*$ to the meson-baryon channels vary quantitatively, as listed in Tables~\ref{table:parameter_ORB}, \ref{table:parameter_OM} and \ref{table:parameter_BNW}.
Reflecting the different properties of $\Lambda^{*}$, the effective $\bar{K}$ mass in Eq.~\ref{eq:effective_mass} and the coupling constants concerning $\Lambda^*$ depend on the model.
By comparing the properties of the $\Lambda^* N$ systems among these chiral unitary models, we may understand mechanisms for the $\Lambda^* N$ potential.

As in the same way with Sec.~\ref{sec:estimation}, we estimate the coupling constants in these models. Values of the estimated coupling constants with the effective $\bar{K}$ masses are listed in Tables~\ref{table:ORB},~\ref{table:OM} and \ref{table:BNW}.
Qualitative features of the parameters of the $\Lambda^{*}N$ potential are almost the same within these models.
We note that the estimated couplings of $\Lambda^*$ to the $\sigma$ and $\omega$ mesons are strongly enhanced, when the mass of $\Lambda^*$ is close to the $\bar{K}N$ threshold, as in the $\Lambda^{*}_{1}$ states in OM and BNW models shown in Tables~\ref{table:OM} and \ref{table:BNW}.
This is caused by the intermediate $\bar{K}N$ loop in Fig.~\ref{fig:estimation}, which becomes large if the intermediate states are close to on their mass shell.
This enhancement, however, does not occur when the width of $\Lambda^*$ is taken into account, so the strong $\Lambda^*$ couplings and the properties of the bound states in these models should be understood with caution.
It is necessary to improve the estimation of the coupling constants, for the consistent treatment of the models in which $\Lambda^*$ lies close to a meson-baryon threshold.
Such refinement is out of the scope of this paper and left for a subject of future works.

\begin{figure}[tb]
  \begin{center}
  \subfigure[$\Lambda^*_1 N,^1S_0$]{
  \includegraphics*[scale=0.7]{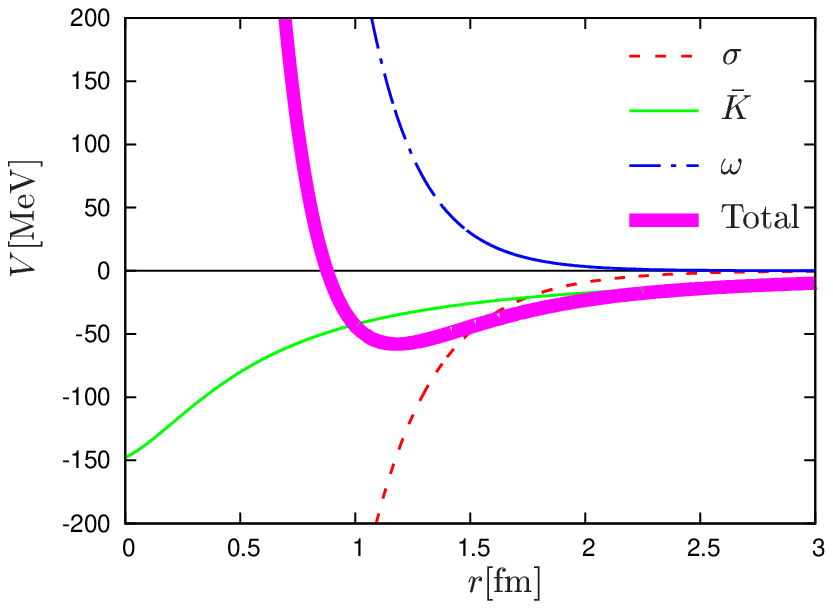}
  \label{fig:V_1_BNW_S0}}
  \subfigure[wave function $\psi(r)$]{
  \includegraphics*[scale=0.7]{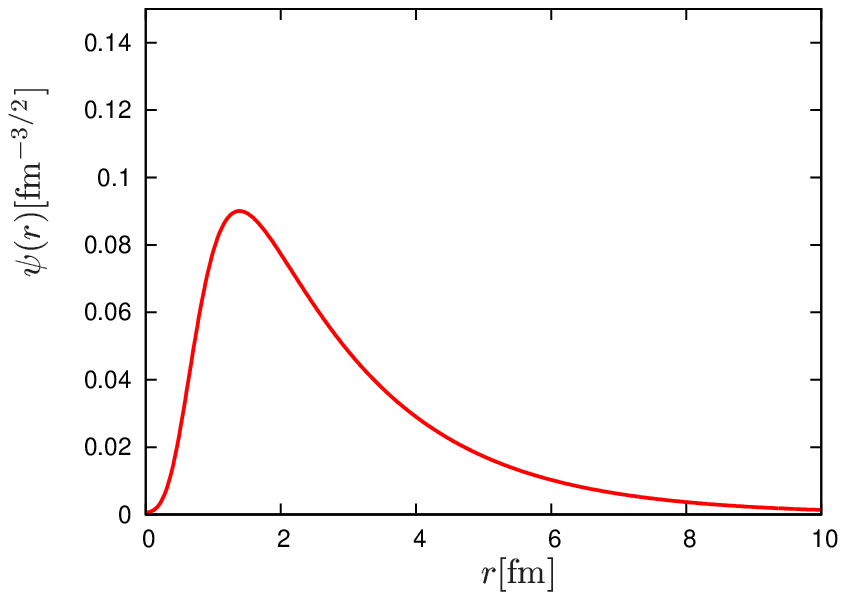}
  \label{fig:wf_BNW}}
  \end{center}
  \caption{The $\Lambda^*_1 N$ component of the potential $V_{11}$ and the wave function of the $\Lambda^*_1 N$ system with the BNW model.}
  \label{fig:BNW}
  \end{figure}

For each $\Lambda^{*}N$ potential with different input, we obtain one quasi-bound state in the $^1S_0$ channel, and the properties of the quasi-bound state are presented in Table~\ref{table:comparison}.
All the quasi-bound states are obtained in a small energy region slightly below the $\bar{K}NN$ threshold. 
However, since the $\Lambda^*_a N$ threshold differs among chiral unitary models, the size of the quasi-bound state is not so close to each other.
For instance, although the mass of the bound state $M_{\Lambda^* N}$ in the HNJH model \cite{Hyodo:2002pk,Hyodo:2003qa} and that of the BNW model \cite{Borasoy:2004kk,Borasoy:2005ie} are much close, the $\Lambda^* N$ system is more compressed in the BNW model than the HNJH model. 
Because the $\Lambda^{*}_{1}N$ threshold in the BNW model is higher than the corresponding threshold in the HNJH model, the two-body $\Lambda^{*}_{1}N$ system should have larger binding energy and hence the smaller size.
For comparison, the $\Lambda^*_1 N$ component of the potential in the $^1S_0$ channel and wave function of the $\Lambda^*_1 N$ bound state are shown in Fig.~\ref{fig:BNW}.

   \begin{table}[tb]
   \caption{Coupling constants and the effective $\bar{K}$ masses of the $\Lambda^*N$ potential in the ORB model\cite{Oset:2001cn}.}
   \begin{center}
   \begin{tabular}{c|rrrrr}
   \hline \hline
   & $g_{\Lambda^* N \bar{K}} / \sqrt{4\pi}$ & $g_{\Lambda^* \Lambda^* \sigma} / \sqrt{4\pi}$ & $g_{\Lambda^* \Lambda^* \omega} / \sqrt{4\pi}$ & $f_{\Lambda^* \Lambda^* \omega} / \sqrt{4\pi}$ & $\tilde{m}_{\bar{K}}$~(MeV) \\ \hline
   $\Lambda^*_1(1426)$ & $0.53$ & 4.43 & 15.71 & 10.31 &  96 \\
   $\Lambda^*_2(1390)$ & $0.41$ & 0.92 &  3.77 &  2.91 & 207 \\
   Transition          & -      & 1.64 &  6.60 &  4.44 & 162 \\
   \hline
   \end{tabular}
   \label{table:ORB}
   \end{center}
   \end{table}
   
   \begin{table}[tb]
   \caption{Coupling constants and the effective $\bar{K}$ masses of the $\Lambda^*N$ potential in the OM model\cite{Oller:2000fj}.}
   \begin{center}
   \begin{tabular}{c|rrrrr}
   \hline \hline
   & $g_{\Lambda^* N \bar{K}} / \sqrt{4\pi}$ & $g_{\Lambda^* \Lambda^* \sigma} / \sqrt{4\pi}$ & $g_{\Lambda^* \Lambda^* \omega} / \sqrt{4\pi}$ & $f_{\Lambda^* \Lambda^* \omega} / \sqrt{4\pi}$ & $\tilde{m}_{\bar{K}}$~(MeV) \\ \hline
   $\Lambda^*_1(1434)$ & 0.37 & 8.33 & 27.32 & 17.82 &  37 \\
   $\Lambda^*_2(1379)$ & 0.22 & 0.39 &  0.97 &  0.81 & 230 \\
   Transition          & -    & 0.79 &  2.41 &  1.51 & 167 \\
   \hline
   \end{tabular}
   \label{table:OM}
   \end{center}
   \end{table}

   \begin{table}[tb]
   \caption{Coupling constants and the effective $\bar{K}$ masses of the $\Lambda^*N$ potential in the BNW model\cite{Borasoy:2004kk,Borasoy:2005ie}.}
   \begin{center}
   \begin{tabular}{c|rrrrr}
   \hline \hline
   & $g_{\Lambda^* N \bar{K}} / \sqrt{4\pi}$ & $g_{\Lambda^* \Lambda^* \sigma} / \sqrt{4\pi}$ & $g_{\Lambda^* \Lambda^* \omega} / \sqrt{4\pi}$ & $f_{\Lambda^* \Lambda^* \omega} / \sqrt{4\pi}$ & $\tilde{m}_{\bar{K}}$~(MeV) \\ \hline
   $\Lambda^*_1(1433)$ & 0.52 & 10.58 & 35.00 & 22.87 &  48 \\
   $\Lambda^*_2(1388)$ & 0.45 &  0.77 &  4.42 &  3.27 & 211 \\
   Transition          & -    &  1.98 &  7.44 &  4.94 & 155 \\
   \hline
   \end{tabular}
   \label{table:BNW}
   \end{center}
   \end{table}

   \begin{table}[tb]
   \caption{The properties of the $\Lambda^* N$ quasi-bound state with several chiral models.
            Each value is obtained by the results of the $\Lambda^*_1 N$ single channel calculation, as discussed in Sec.~\ref{sec:coupled}.
            The binding energy $B$ is measured from the $\Lambda^*_1 N$ threshold.
           }
   \begin{center}
   \begin{tabular}{l|rrr}
   \hline \hline
   Models.                                  & $M_{\Lambda^* N}$~(MeV) & $B$~(MeV) & $\sqrt{\langle r^2 \rangle}$~(fm) \\ \hline
   HNJH\cite{Hyodo:2002pk,Hyodo:2003qa}     & $2365$ & $1.0$ & $5.7$   \\
   ORB\cite{Oset:2001cn}                    & $2364$ & $0.5$ & $6.8$   \\
   OM\cite{Oller:2000fj}                    & $2371$ & $1.8$ & $5.8$   \\
   BNW\cite{Borasoy:2004kk,Borasoy:2005ie}  & $2366$ & $5.9$ & $3.6$   \\
   \hline
   \end{tabular}
   \label{table:comparison}
   \end{center}
   \end{table}
   
\subsection{Dependence on the form factor }

We have introduced the ratio $c$ in Eq.~\ref{eq:c} in order to take into account the difference of the size of $\Lambda^*$ and the nucleon.
Here we consider the effect of the size of $\Lambda^*$ to the bound state, by varying the ratio $c$.
Based on the evaluation of the electromagnetic properties of $\Lambda^*$ in the chiral unitary approach\cite{Sekihara:2008qk,Sekihara:2010uz}, we have used the value $c=1.5$ for both $\Lambda^*_1$ and $\Lambda^*_2$ so far.
If $\Lambda^*$ is a meson-baryon molecule state, the size is expected to be larger than the nucleon.
A large value for $c$ stands for the loose bound of $\Lambda^*$, and leads to the small cut-off for the $\Lambda^* \Lambda^* X$ vertices.
On the other hand, it follows from Eq.~\ref{eq:form_factor} that the cutoff $\Lambda$ should be larger than the mass of the exchanged meson. For instance, if $c$ is larger than 1.92, then $\Lambda_{\Lambda^* \Lambda^* \omega}$ becomes smaller than the mass of the omega meson, which should be avoided in the physical situation.
Therefore the parameter $c$ cannot be arbitrarily large and has an upper limit.
Although the size of $\Lambda^*$ is in principle related to the properties of $\Lambda^*$, coupling strengths to each meson-baryon channel and pole positions, in this section, we simply vary the ratio $c$ from 1 to 1.5 keeping the other parameters unchanged.

\begin{figure}[tb]
  \begin{center}
  \subfigure[Two-body mass.]{
  \includegraphics*[scale=0.7]{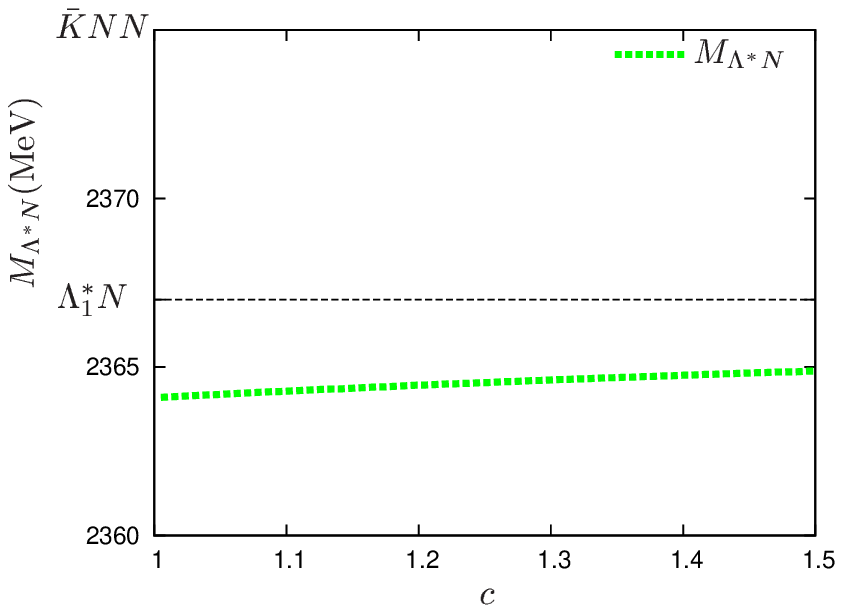}
  \label{fig:c_be}}
  \subfigure[Mean distance.
            ]{
  \includegraphics*[scale=0.7]{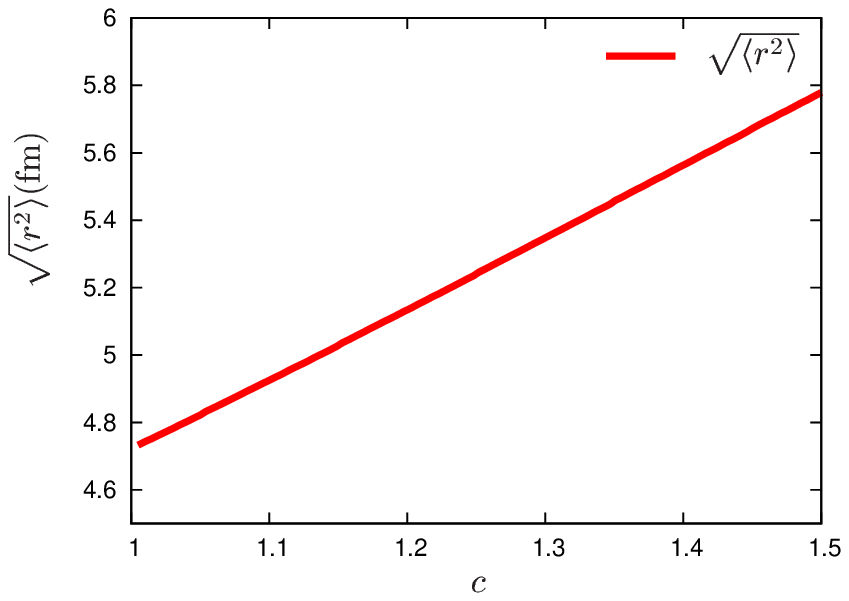}
  \label{fig:c_rms}}
  \end{center}
  \caption{The $c$ dependence of the mass and mean distance of the quasi-bound states of the $\Lambda^* N$.}
  \label{fig:c}
  \end{figure}

In the same manner as Sec.~\ref{sec:single}, we obtain the two-body mass and the mean distance of the $\Lambda^{*}_{1}N$ bound state as functions of the parameter $c$ shown in Fig.~\ref{fig:c}.
It can be seen that the small value for $c$ generates the $\Lambda^*_1 N$ bound state with a little deeper binding and smaller spatial size.
Although the two-body mass does not depend on the ratio $c$ so much, the size of the quasi-bound state becomes small as we decrease the ratio $c$ because the size is sensitive to the binding energy measured from the $\Lambda^*_1 N$ threshold.

\begin{figure}[tb]
  \begin{center}
  \subfigure[$c=1.1$]{
  \includegraphics*[scale=0.7]{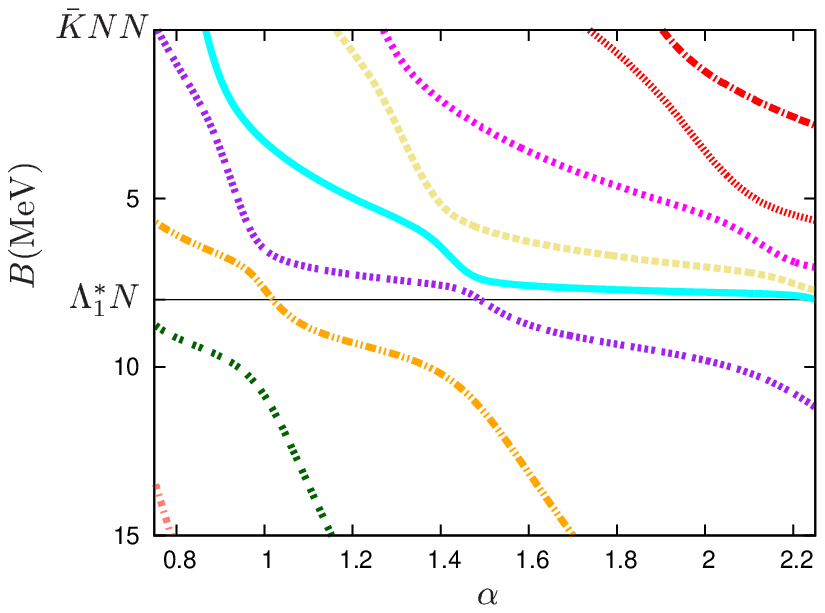}
  \label{fig:rs_c11}}
  \subfigure[$c=1.3$]{
  \includegraphics*[scale=0.7]{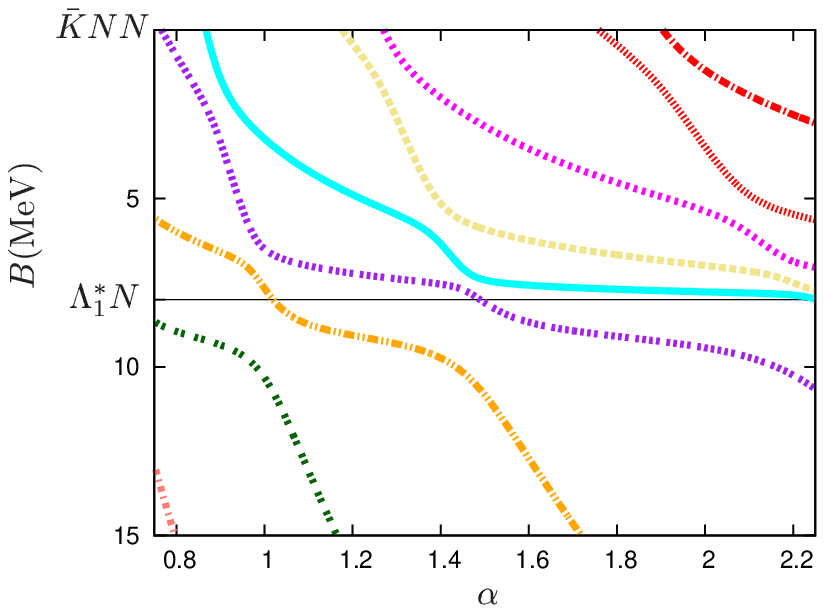}
  \label{fig:rs_c13}}
  \end{center}
  \caption{Energy eigenvalues of the $\Lambda^* N$ system with the HNJH model\cite{Hyodo:2002pk,Hyodo:2003qa} obtained by the real scaling method, for $c=1.1$ and $c=1.3$.
          The parameter $\alpha$ controls the range of the wave function.}
  \label{fig:rs_c}
  \end{figure}

We next show the results of the full channel coupling for the $c=1.1$ and $c=1.3$ case in Fig.~\ref{fig:rs_c}.
In each case, we have one quasi-bound state, whose binding energy from the $\bar{K}NN$ threshold is about 9 MeV.
Because the mass shift from the single-channel result is small in the region $1 \leq c \leq 1.5$, we expect that the $\Lambda^*_1 N$ component of the quasi-bound state is dominant.
The decay width of the process $\Lambda^*_1 N \to \Lambda^*_2 N$, which can be roughly estimated by the distance at the level crossing point, increases when the parameter $c$ is small.
In other words, the $\Lambda^* N$ channel mixing effect becomes larger as the size of $\Lambda^*$ gets smaller.

\subsection{The coupling of the exchanged meson to the pseudoscalar meson}
\label{sec:}

To evaluate the vertex function $\Gamma_{\Lambda^* \Lambda^* X}^M$ in Eq.~\ref{eq:estimation}, the coupling constants of the exchanged mesons to the pseudoscalar meson in the meson-baryon multiple scattering, namely the $\sigma \pi \pi$, $\sigma K \bar{K}$ and $\omega K \bar{K}$ coupling constants, are needed.
In determination of these values, we have notable ambiguities in two points.

One is the $\sigma K \bar{K}$ coupling constant.
The $\sigma \pi \pi$ coupling can be determined by its decay, while the strength of the $\sigma K \bar{K}$ coupling depends on the theoretical models.
We mainly adopt the $\sigma K \bar{K}$ coupling as one half of the $\sigma \pi \pi$ coupling, following a recent determination in the analysis of $\pi \pi$ scattering~\cite{Kaminski:2009qg}.
Whereas the $\sigma K \bar{K}$ coupling may be much smaller than the $\sigma \pi \pi$ coupling, based on the earlier investigations~\cite{Ishida:1995xx,Oller:1998zr}.
Therefore, we study the $\sigma K \bar{K}$ coupling dependence, changing the ratio $r_{\pi K}$ between the couplings, $g_{\sigma K \bar{K}}$ and $g_{\sigma \pi \pi}$
   \begin{eqnarray}
   r_{\pi K}
   &=&
   \frac{ g_{\sigma K \bar{K}} }{ g_{ \sigma \pi \pi} } .
   \end{eqnarray}
We vary $r_{\pi K}$ : from 0 to 1/2.

The other is the normalization of the couplings.
In general, the strength of the coupling constant depends on the momentum transfer.
In order to estimate the coupling of $\Lambda^*$ and the isoscalar meson, we take the soft limit for the exchanged meson in the Breit frame, and thus we need the coupling constant determined under the condition where the four dimensional momentum of the exchanged meson $k$ is taken as $k=(0, \vec{0})$.
On the other hand, the coupling constant $g_{\sigma \pi \pi}$ is determined by the on-shell kinematics of the decay process.
For the case that the exchanged meson couples to the baryon, the momentum dependence is taken into account by the form factor $F_{\alpha}(\vec{k})$ in Eq.~\ref{eq:form_factor} and we renormalize the coupling constant by multiplying $F_{\alpha}(0)$.
Although we do not have the cut-off mass for the vertices where three mesons couple, with reference to $F_{\alpha}(0) : 0.7 \sim 0.9$, we can estimate the effect of the normalization of the couplings by introducing the same factor $\beta$ for all vertices as
   \begin{eqnarray}
   g_{\sigma \pi \pi } , g_{\omega K \bar{K} }
   &\to&
   \beta g_{\sigma \pi \pi } , \beta g_{\omega K \bar{K} } ,
   \end{eqnarray}
where $\beta$ is moved from 0.7 to 1.0.
The $\sigma K \bar{K}$ coupling automatically changes with the $\sigma \pi \pi$ coupling, while the ratio $r_{\pi K}$ is fixed to be $1/2$.

Considering the above ambiguities, we have found that the qualitative features do not change.
In terms of the binding energy, the deviation due to the ambiguities is around 1 MeV.
Thus, we conclude that our results are not sensitive to the ambiguities in the meson coupling constant.

\subsection{The binding mechanism in the $\Lambda^{*}N$ system and comparison with other works}
  
In summary of the discussion of the theoretical uncertainties, we  conclude that the $\Lambda^*N$ system forms a bound state in a wide range of the parameter space. 
The quantitative results of the mass and wave function, however, depend on the input of the $\Lambda^*N$ potential. 
The relation between the $\Lambda^*$ mass and the binding energy of the $\Lambda^*N$ is worth mentioning. 
In the $\bar{K}N$ bound state picture for $\Lambda^*$, the shallow binding of $\Lambda^*$ leads to the spatially large size, and effective $\bar{K}$ mass in the $\Lambda^*N$ potential should be small. 
Both effects enhance the $\bar{K}$ exchange contribution in the $\Lambda^*N$ potential, and hence the binding energy of the $\Lambda^*N$ system increases. 
Thus, in contrast to the naive expectation, the small binding energy of $\Lambda^*$ in the $\bar{K}N$ system leads to the larger binding of the $\Lambda^*N$ system in the $\Lambda^*N$ potential picture.
To pin down the precise position of the $\Lambda^*N$ bound state, we need further understanding of the properties of $\Lambda^*$.

We compare the present result with other theoretical works for the $\bar{K}NN$ system. 
The main difference from the previous study of the $\Lambda^{*}N$ potential in Ref.~\cite{Arai:2007qj} is the binding mechanism. 
In Ref.~\cite{Arai:2007qj}, the potential was constructed to support a bound state with 88 MeV below the threshold by adjusting the $\Lambda^{*}\Lambda^{*}\sigma$ coupling constant. 
The bound state was generated by the short range attraction, and hence the wave function was very much compressed. 
The contribution from the $\bar{K}$ exchange was very small, due to the small $\Lambda^{*}N\bar{K}$ coupling. 
On the other hand, present $\Lambda^{*}N$ potential is constructed based on the microscopic description of $\Lambda^{*}$ in the chiral unitary approach, and the binding energy of the $\Lambda^{*}N$ system is a prediction of the model. 
Because of the strong $\Lambda^{*}N\bar{K}$ coupling discussed in Sec.~\ref{sec:K}, the $\bar{K}$ exchange diagram plays a major role to generate a bound state in the $S=0$ channel. 
Due to the lighter effective $\bar{K}$ mass, the attraction has longer range. In the end, we obtain a loosely bound $\Lambda^{*}N$ system which seems to be more compatible with the meson-exchange potential picture.

The $\Lambda^{*}N$ bound state appears at $\sim$ 10 MeV below the $\bar{K}NN$ threshold. 
This is in the same line with the results which utilize the chiral SU(3) dynamics to constrain the meson-baryon interaction; the three-body variational calculation with effective $\bar{K}N$ interaction\cite{Dote:2008in,Dote:2008hw},  the coupled-channel Faddeev calculation with energy-dependent interaction~\cite{Ikeda:2010tk}, and the fixed-center approximation to the Faddeev approach for $\bar{K}NN$ system\cite{Bayar:2011qj}. 
It is worth noting that Ref.~\cite{Ikeda:2010tk} found two poles in the $\bar{K}NN$-$\pi\Sigma N$ amplitude. 
One pole corresponds to the shallow bound state obtained in our model, while there is another state with larger binding energy with huge width of $\sim 244-320$ MeV. 
This state may be related with the lower energy $\Lambda^{*}_{2}N$ state.


\section{Conclusion}
\label{sec:conclusion}

We study the bound state solution of $\Lambda(1405)$ and a nucleon ($\Lambda^* N$) system which is the simplest $\Lambda^{*}$ hypernucleus.
We examine the interaction of the $s$-wave $\Lambda^{*}N$ system with the total spin $S=0$ and $S=1$.
We construct the one-boson-exchange $\Lambda^* N$ potential by extending the J\"{u}lich potential.
Exchanges of $\sigma$, $\omega$ and $\bar{K}$ mesons are considered and their couplings to the $\Lambda^*$ baryon are evaluated from the properties of $\Lambda^*$ in the chiral unitary approach.
Reflecting the two-pole picture of $\Lambda^*$ in the chiral unitary approach, the $\Lambda^* N$ two-body system should consist of two components.
We call the higher energy state as $\Lambda^*_1$ and the lower energy state as $\Lambda^*_2$. 
The one-boson-exchange potential allows the transition from the $\Lambda^{*}_{1}N$ state to the $\Lambda^{*}_{2}N$ state and \textit{vice versa}.

In the chiral unitary approach, $\Lambda^*$ is described as a quasi-bound state of the $\bar{K}N$ system, so the $\bar{K}$ exchange contribution to the $\Lambda^*N$ potential plays an important role.
It is found that the $\Lambda^* N \bar{K}$ coupling constant is much stronger than the value estimated by the decay width of $\Lambda^*$ and the SU(3) symmetry.
The $\bar{K}$ exchange is attractive (repulsive) in the spin $S=0$ ($S=1$) channel, and this contribution dominates the volume integral of the potential due to the light effective $\bar{K}$ mass.

In the single channel calculation which does not include the channel mixing, the $\Lambda^*_1 N$ system in the $^1S_0$ channel has a bound state with the mass $M_{\Lambda^* N} =$ 2365 MeV slightly below the threshold.
Considering the mixing effect, we have one quasi-bound state with a finite decay width.
The energy shift due to the mixing is small, and thus the $\Lambda^* N$ quasi-bound state can be considered as being dominated by the $\Lambda^*_1 N$ component.
The internal structure of the $\Lambda^* N$ quasi-bound state can be extracted by the use of the wave function of the $\Lambda^*_1 N$ component.
The mean distance of the $\Lambda^{*}N$ system is $\sqrt{\langle r^2 \rangle} \sim$ 5.7 fm.
In our present model, the $\Lambda^* N$ quasi-bound state is found to be a loosely bound system.

The $\Lambda^*N$ potential model treats $\Lambda^*$ as a fundamental particle, while it has finite decay width in vacuum.
Thus, the $\Lambda^* N$ bound state in this study has various decay modes, namely, the non-mesonic decays into $\Lambda N$ and $\Sigma N$ channels and the mesonic decay modes of the $\pi\Sigma N$ and the $\pi\Lambda N$.
These decays can be studied by combining the wave function obtained in this paper with the transition amplitude of the decay process as studied in Ref.~\cite{Sekihara:2009yk}.
Such a study is underway.

In the present work, we have constructed the two-body bare $\Lambda^* N$ potential in vacuum, which is the fundamental building block in the $\Lambda^*$-hypernuclei picture.
The few body $\Lambda^*$-hypernuclei, like the $\Lambda^* NN$ three-body system, can be studied by the wisdom of the few-body technique, developed for the normal nuclei and hypernuclei~\cite{Hiyama:2010zzb,Hiyama:2003cu}. 
The effective $\Lambda^* N$ interaction in nuclear matter may be constructed by the G-matrix method.
Thus, the $\Lambda^{*}N$ potential constructed in the present work will bring new perspective of the $\Lambda^*$-hypernuclei to the physics of the strangeness nuclear physics.


\section*{Acknowledgements}

We thank Drs. W. Weise, A. Gal, E. Hiyama and Y. Ikeda for useful discussion.
This work is partially supported by a grant for the Tokyo Institute of Technology Global COE program, ``Nanoscience and Quantum Physics", from the Ministry of Education, Culture, Sports, Science and Technology of Japan, and by the Grant-in-Aid for Scientific Research from 
MEXT and JSPS (Nos.
  19540275, 
  21840026, 
  and 22105503).


\appendix


\section{$\Lambda^*N$ one-boson-exchange potential}
\label{sec:obep}

The explicit forms of the $\sigma,\omega$ and $\bar{K}$ exchange contributions to the $\Lambda^{*}N$ potential, $V_{\sigma},V_{\omega}$ and $V_{\bar{K}}$ in Eq.~\ref{eq:lambda_potential} are given in this Appendix.
The necessary parameters to construct the $\Lambda^* N$ potential, $i.e.$, the coupling constants and cut-off masses, are listed in Table~\ref{table:g_HNJH} and Table~\ref{table:vertex_Juelich}.
These one-boson-exchange potentials are derived by the standard manner, following the J\"{u}lich model\cite{Holzenkamp:1989tq,Reuber:1992dh,Reuber:1993ip}. 
The leading order terms in the static approximations of baryons, which are relevant to the $s$-wave state, are given by 
   \begin{eqnarray}
   V_\sigma(m_\sigma,r)&=&-\frac{g_{\Lambda^*\Lambda^*\sigma}g_{NN\sigma}}{4\pi}m_{\sigma}\left(
                                                                           1
                                                                           -\frac{m_{\sigma}^2}{8M_{\Lambda^*}M_N}
                                                                           \right)\phi(m_\sigma r) \label{eq:sigma_potential} \ ,\\ 
   V_\omega(m_\omega,r)&=&\frac{m_{\omega}}{4\pi}\left[
                                  \left\{
                                  g_{\Lambda^*\Lambda^*\omega}g_{NN\omega}\left(
                                              1+\frac{m_{\omega}^2}{8M_{\Lambda^*}M_N}
                                              \right)
                                  +g_{\Lambda^*\Lambda^*\omega}f_{NN\omega}\frac{m_{\omega}^2}{4{\cal M}M_N}
                                  \right.
                                  \right.
                                  \nonumber \\
                                  &&\left.
                                  +g_{NN\omega}f_{\Lambda^*\Lambda^*\omega}\frac{m_{\omega}^2}{4{\cal M}M_\Lambda^*}
                                  \right\}
                                  \nonumber \\
        &&+\frac{m_{\omega}^2}{4M_\Lambda^*M_N}\frac{2(\vec{\sigma}_1\cdot\vec{\sigma}_2)}{3} \nonumber \\
        &&\times \left\{
                 g_{\Lambda^*\Lambda^*\omega}g_{NN\omega}
                 +g_{\Lambda^*\Lambda^*\omega}f_{NN\omega}\frac{M_N}{{\cal M}}
                 +f_{\Lambda^*\Lambda^*\omega}g_{NN\omega}\frac{M_\Lambda^*}{{\cal M}}
                 \right.
                 \nonumber \\
                 &&
                 \left.
                 \left.
                 +f_{\Lambda^*\Lambda^*\omega}f_{NN\omega}\frac{M_\Lambda^*M_N}{{\cal M}^2}
                 \right\}
                 \right]\phi(m_\omega r) \label{eq:omega_potential} \ ,\\
   V_{\bar{K}}(\tilde{m}_{\bar{K}},r)&=&\left\{
              \frac{1+(\vec{\sigma}_{1}\cdot\vec{\sigma}_{2})}{2}
              \right\}\frac{g_{\Lambda^*N\bar{K}}g_{\Lambda^*N\bar{K}}}{4\pi} \nonumber\\
              &&\times \tilde{m}_{\bar{K}}
                                               \left(
                                               1
                                               -\frac{\tilde{m}_{\bar{K}}^2}{8M_{\Lambda^*}M_N}
                                               \right)\phi(\tilde{m}_{\bar{K}} r) \label{eq:kbar_potential} \ ,
   \end{eqnarray}
with
   \begin{eqnarray}
   \phi(x)&=&\frac{e^{-x}}{x}
   \ ,
   \end{eqnarray}
where ${\cal M}$ is the scaling mass chosen to be the proton mass and $\vec{\sigma}_i$ denotes the spin operator of the baryon $i$.
In the $\bar{K}$ exchange potential (\ref{eq:kbar_potential}), the exchange factor in Eq.~\ref{eq:spin_exchange} is included and effective $\bar{K}$ mass in Eq.~\ref{eq:effective_mass} is used as noted in Sec.~\ref{sec:K}.
  
   \begin{table}[tb]
   \caption{Coupling constants and the cutoff parameters in the J\"{u}lich model.}
   \begin{center}
   \begin{tabular}{c|ccc}
   \hline \hline
   vertex & $g/\sqrt{4\pi}$ & $f/\sqrt{4\pi}$ & $\Lambda$(GeV) \\ \hline
   $\Sigma \Sigma \sigma$ & 3.061 &   -   & 1.0 \\
   $NN\sigma$             & 2.385 &   -   & 1.7 \\
   $\Sigma \Sigma \omega$ & 2.981 & 2.796 & 2.0 \\
   $NN\omega$             & 4.472 &   0   & 1.5 \\
   $NN\pi$                & 3.795 &   -   & 1.3 \\ \hline
   \end{tabular}
   \label{table:vertex_Juelich}
   \end{center}
   \end{table}

In the derivation of the potential in the momentum space, following higher momentum $\vec{k}$ term on intermediate mesons appears
   \begin{eqnarray}
   \frac{\vec{k}^2}{\vec{k}^2+m^2}=1-\frac{m^2}{\vec{k}^2+m^2}
   \ .
   \end{eqnarray}
The first term of the right hand side leads to the $\delta$-function in the coordinate space by Fourier transformation, which we omit, following the J\"{u}lich model\cite{Holzenkamp:1989tq}. 
One sees that the $\omega$ and $\bar{K}$ exchange contributions depend on the total spin $S$, while the $\sigma$ exchange term is spin independent, from Eqs.~\ref{eq:sigma_potential}, \ref{eq:omega_potential} and \ref{eq:kbar_potential}.


\section{$\Lambda^*$ coupling constants}
\label{sec:coupling}

As noted in Sec.~\ref{sec:estimation}, the $\Lambda^*$ coupling constants are estimated and numerical results in the HNJH model are listed in Table~\ref{table:g_HNJH}.
We show how we estimate these coupling constants on the $\Lambda^*$ chiral unitary approach, in detail.
We calculate the one-loop diagrams shown in Fig.~\ref{fig:estimation}, based on the meson-baryon molecule picture of $\Lambda^*$\cite{Hyodo:2008xr}.
In the study of the electromagnetic form factors in Refs.~\cite{Sekihara:2008qk,Sekihara:2010uz}, it can be shown that this method to evaluate the coupling constant is exact on top of the pole position.
The interaction Lagrangian for $\tilde{\Gamma}^{(i)}_{\Lambda^* BM}$, where the index $i=1(i=2)$ represents the $\pi\Sigma (\bar{K}N)$ channel, is given as a scalar type
   \begin{eqnarray}
   {\cal L}^{int}=g_{\Lambda^*_a BM}^{(1)}\bar{\Lambda^*_a}
   [\pi\Sigma]_{I=0} 
   + g_{\Lambda^*_a BM}^{(2)}\bar{\Lambda^*_a}
   [\bar{K}N]_{I=0}
   +h.c.
   \label{eq:chiral_Lagrangian}
   \ ,
   \end{eqnarray}
where $g_{\Lambda^*_{a} B M}^{(i)}$ is the coupling constant of $\Lambda^*_{a}$ to the channel $i$ which are listed in Table~\ref{table:parameter_HNJH}.
\footnote{
In this appendix, we use the $\Lambda^* BM$ coupling constants in the isospin basis (\ref{eq:chiral_Lagrangian}). 
Since $\Lambda^*$ is isospin zero, the $\Lambda^* BM$ coupling constants in the charge basis have additional factor $1/\sqrt{N}$ with the isospin degeneracy $N$, as in Eq.~\ref{eq:coupling}. 
This factor is cancelled by the summation over intermediate states in (\ref{eq:est_B}) and (\ref{eq:est_M}). 
Thus, the final result remains unchanged in the charge basis. 
For other vertices, we use the particle basis for the coupling constants, $g_{NN\sigma}=g_{pp\sigma}=g_{nn\sigma}$, and so on.
}
This leads to the vertex
   \begin{eqnarray}
   \tilde{\Gamma}_{\Lambda^*_a BM}^{(i)}=g_{\Lambda^*_a BM}^{(i)}I
   \ ,
   \end{eqnarray}
where $I$ represents the unit matrix in the spinor space.

For the evaluation of the coupling constants, we take the Breit frame where the momentum $\vec{p}_a$ ($\vec{p}_b$) of the initial (final) baryon is given by $-\vec{k}/2$ ($\vec{k}/2$) and the energy transfer $k^0$ is zero
   \begin{eqnarray}
   k&=&\left(
          0,\vec{k}
     \right)
     \ , \\
   p_a&=&\left(
         \sqrt{M_{\Lambda^*_a}^2+\frac{\vec{k}^2}{4}},-\frac{\vec{k}}{2}
         \right) \ ,\\
   p_b&=&\left(
         \sqrt{M_{\Lambda^*_b}^2+\frac{\vec{k}^2}{4}},\frac{\vec{k}}{2}
         \right) \ .
   \end{eqnarray}
Considering that the vertices $\Gamma_{\Lambda^* \Lambda^* X}$ have the indices $ba$ corresponds to each matrix element of the $\Lambda^* N$ potential and depend only on the momentum transfer $\vec{k}$ in this frame, Eq.~\ref{eq:estimation} can be rewritten as
   \begin{eqnarray}
   \left( \Gamma_{\Lambda^* \Lambda^* X} \right)_{ba}(\vec{k})=\left( \Gamma_{\Lambda^* \Lambda^* X}^B \right)_{ba} (\vec{k}) + \left( \Gamma_{\Lambda^* \Lambda^* X}^M \right)_{ba} (\vec{k}) \ ,
   \end{eqnarray}
   with
   \begin{eqnarray}
   \left(
   \Gamma_{\Lambda^* \Lambda^* X}^B
   \right)_{ba}
   (\vec{k})
   &=&
   \sum_{i=1,2}
   \left(
   g_{\Lambda^* BM}^{(i)}
   \right)^2_{ba}
   \int\frac{d^4 q}{i(2\pi)^4}
   \frac{1}{m_i^2-q^2}\frac{1}{M_i-(\dslash{p}_b-\dslash{q})}
   \nonumber \\*
   &&\times
   \tilde{\Gamma}_{BBX}^{(i)}(\vec{k})\frac{1}{M_i-(\dslash{p}_a-\dslash{q})}
   \label{eq:est_B} \ , \\
   \left(
   \Gamma_{\Lambda^* \Lambda^* X}^M
   \right)_{ba}
   (\vec{k})
   &=&
   \sum_{i=1,2}
   \left(
   g_{\Lambda^* BM}^{(i)}
   \right)^2_{ba}
   \int\frac{d^4 q}{i(2\pi)^4}
   \frac{1}{M_i-\dslash{q}}\frac{1}{m_i^2-(p_b-q)^2} \nonumber \\
   && \times
   \tilde{\Gamma}_{MMX}^{(i)}(\vec{k})\frac{1}{m_i^2-(p_a-q)^2}
   \label{eq:est_M} \ ,
   \end{eqnarray}
where $m_i$ and $M_i$ represent masses of the meson and the baryon in channel $i$, and
   \begin{eqnarray}
   \left(
   g_{\Lambda^* BM}^{(i)}
   \right)^2_{ba}
   =g_{\Lambda^*_b BM}^{(i)} g_{\Lambda^*_a BM}^{(i)}.
   \label{eq:est_g^2}
   \end{eqnarray}
$\tilde{\Gamma}_{NNX}^{(i)}$ and $\tilde{\Gamma}_{MMX}^{(i)}$ correspond to the vertices where the isoscalar meson $X$ couples to the intermediate baryons and mesons in channel $i$.
In this way, we obtain the momentum dependent vertex $\left( \Gamma_{\Lambda^* \Lambda^* X}\right)_{ba} (\vec{k})$.
Since the form of $\tilde{\Gamma}_{BBX}^{(i)}$ and $\tilde{\Gamma}_{MMX}^{(i)}$ depend on the meson $X$, in the followings, we separately discuss the explicit form of vertices for the $\sigma$ and $\omega$ exchanges.
The properties of $\Lambda^*$ obtained by the HNJH chiral unitary model\cite{Hyodo:2002pk,Hyodo:2003qa} are listed in Table~\ref{table:parameter_HNJH}, while the other model cases are discussed in Sec.~\ref{sec:dis_chiral}.

In order to translate the resulting vertices into the $\Lambda^* N$ potential model, we estimate the coupling strength by taking the soft limit $\vec{k} \to \vec{0}$, while the momentum dependence of the coupling is modeled by the phenomenological form factor, as shown in Sec.~\ref{sec:Form factor}.
It is in principle possible to use the $\vec{k}$ dependence of the vertex function as the form factor in the potential model, but it leads to a very complicated form of the potential in $r$ space.
At the present stage, the size of $\Lambda^*$ (and hence the magnitude of the cutoff) is more relevant quantity to the result, rather than the detailed structure of the momentum dependence.
Thus, it is sufficient to adopt the phenomenological form factor with the size of $\Lambda^*$ estimated in the same framework~\cite{Sekihara:2008qk,Sekihara:2010uz}.

In the following section, we omit the indices $ba$ of the $\Lambda^* \Lambda^* X$ vertices for simplicity.
In the soft limit, parameters which have the index $a,b$ is $g_{\Lambda^* BM}^{(i)}$ and $M_{\Lambda^*}$ in Eqs.~\ref{eq:est_B} and \ref{eq:est_M}.
For the coupling $g_{\Lambda^* BM^{(i)}}$, we follow Eq.~\ref{eq:est_g^2}. The $\Lambda^*$ mass in the off-diagonal component is defined as
   \begin{eqnarray}
   M_{\Lambda^*}=\frac{M_{\Lambda^*_1}+M_{\Lambda^*_2}}{2}.
   \end{eqnarray}


\subsection{$\Lambda^*\Lambda^*\sigma$ coupling}

The $\sigma$ meson couples to both the baryon and meson in the multiple scattering.
Accordingly, we show the way to calculate each contribution, and then, combine them to obtain the $\Lambda^* \Lambda^* \sigma$ coupling.
The interaction Lagrangian between the $\sigma$ meson and the baryon $\Sigma$ and $N$ in the multiple scattering, can be written as a scalar type
   \begin{eqnarray}
   {\cal L}^{int}=g_{BB\sigma}^{(1)} \bar{\Sigma} \Sigma \sigma + g_{BB\sigma}^{(2)} \bar{N} N \sigma,
   \label{eq:Lag_psipsisig}
   \end{eqnarray}
and the coupling constants are given by the J\"{u}lich model, listed in Table~\ref{table:vertex_Juelich}.
In the same manner as Sec.~\ref{sec:Form factor}, we include the momentum dependence in the vertices $\tilde{\Gamma}_{BBX}$ by the use of the phenomenological form factor $F(\vec{k})$ in Eq.~\ref{eq:form_factor}.
Then, we have the vertices $\tilde{\Gamma}_{BB\sigma}^{(i)}(\vec{k})$ with the coupling constant $g_{BB\sigma}^{(i)}$, given by
   \begin{eqnarray}
   \tilde{\Gamma}_{BB\sigma}^{(i)}(\vec{k})&=&g_{BB\sigma}^{(i)}F_{BB\sigma}^{(i)}(\vec{k})I
   \label{eq:tilGamma_Jsig}
   \ .
   \end{eqnarray}
By substituting Eq.~\ref{eq:tilGamma_Jsig} into Eq.~\ref{eq:est_B} with $X=\sigma$, we obtain
   \begin{eqnarray}
   \Gamma_{\Lambda^* \Lambda^* \sigma}^B(\vec{k})
   & \to &
   \Gamma_{\Lambda^* \Lambda^* \sigma}^B(0) \nonumber \\
   &=&
   \sum_{i=1,2} \left(g_{\Lambda^* BM}^{(i)} \right)^2 \left( I_{\sigma,B0}^{(i)} I + I_{\sigma,B1}^{(i)} \gamma^0 \right)
   \label{eq:Gamma_EsigB}
   \ ,
   \end{eqnarray}
with
   \begin{eqnarray}
   I_{\sigma,B0}^{(i)}
   &=&
   \frac{ g_{BB\sigma}^{(i)}F_{BB\sigma}^{(i)}(0) }{(4\pi)^2} \nonumber \\
   &\times&
   \int_0^1 y
   \left[
   \frac{ M_{\Lambda^*}^2 (1-y)^2 + M_i^2 }{ h_B^{(i)}(y) }
   \right.\nonumber \\
   &&\left.
      +2\left\{
         a_i(\mu) + \ln \left( \frac{ h_B^{(i)}(y) }{\mu^2} \right)
         \right\}
      \right] dy \ ,
      \label{eq:divloop}\\
   I_{\sigma,B1}^{(i)}
   &=&
   \frac{ g_{BB\sigma}^{(i)}F_{BB\sigma}^{(i)}(0) }{(4\pi)^2}
   \int_0^1 
   \frac{2M_{\Lambda^*}M_iy} { h_B^{(i)}(y) } dy \ ,
   \end{eqnarray}
where
   \begin{eqnarray}
   h_B^{(i)}(y)
   &=&
   y(M_i^2-M_{\Lambda^*}^2)+y^2M_{\Lambda^*}^2+(1-y)m_i^2\nonumber \\
   &=&
   M_{\Lambda^*}^2y^2+(M_i^2-M_{\Lambda^*}^2-m_i^2)y+m_i^2
   \label{eq:m^2_B+p^2_B}
   \ .
   \end{eqnarray}
The term proportional to $\gamma^0$ come from our choice of the Breit frame and it is associated with the zeroth component of the $\Lambda^*$ momentum, which is $M_{\Lambda^*}$ in the $\vec{k}=0$ limit.
In the calculation of \ref{eq:divloop}, we performed the dimensional regularization to tame the divergence of the loop integral, using the same subtraction scale $\mu$ and the subtraction constant $a(\mu)$ as the model of the dynamical generation of $\Lambda^*$, which are listed in Table~\ref{table:parameter_HNJH}. The other loop diagram converges and no regularization parameter is needed.

For remaining contribution, the interaction Lagrangian between the $\sigma$ meson and two mesons $\pi\pi$ or $K\bar{K}$ are defined as
   \begin{eqnarray}
   {\cal L}^{int}=g_{MM\sigma}^{(1)}\sigma \vec{\pi} \cdot \vec{\pi} + g_{MM\sigma}^{(2)}\sigma \bf{K} \cdot \bar{\bf{K}}
   \ ,
   \end{eqnarray}
with isotriplet pion and isodoublet kaon fields
   \begin{eqnarray}
   \vec{\pi}&=&(\pi_1,\pi_2,\pi_3)\ , \\
   \bf{K}&=&(K^+,K^0)
   \ .
   \end{eqnarray}
The $\sigma \pi \pi$ coupling constant can be determined by the $\sigma$ decay, while the $\sigma K \bar{K}$ coupling is not determined experimentally because the $K \bar{K}$ threshold is above the $\sigma$ mass.
However, the $\sigma K \bar{K}$ coupling is discussed with various theoretical approaches~\cite{Kaminski:2009qg,Ishida:1995xx,Oller:1998zr}.
Here we follow the recent study~\cite{Kaminski:2009qg} and the $\sigma K \bar{K}$ coupling is assumed to be one half of $g_{\sigma \pi \pi}$.
Although the $\sigma \pi \pi$ coupling constant should also be renormalized with the form factor, we have no information of the cut-off for the $\sigma \pi \pi$ and $\sigma K \bar{K}$ vertices and the $\sigma$ meson has ambiguities for its mass and width.
Then, based on the several works, we determine the $\sigma \pi \pi$ coupling with the condition that the mass of $\sigma$ is 550 MeV which is used in the J\"{u}lich model.
So the vertices $\tilde{\Gamma}_{MM\sigma}^{(i)}$ in the soft limit are given by
   \begin{eqnarray}
   \tilde{\Gamma}_{MM\sigma}^{(i)}|_{\vec{k}=0}
   &=&
   g_{MM\sigma}^{(i)}
   \ ,
   \end{eqnarray}
where the coupling constants are chosen to be
   \begin{eqnarray}
   g_{MM \sigma}^{(1)}
   &=&
   2600~{\rm MeV}, \\
   g_{MM \sigma}^{(2)}
   &=&
   \frac{g_{MM \sigma}^{(1)}}{2}.
   \end{eqnarray}
Accordingly, the vertex function in the soft limit can be estimated as
   \begin{eqnarray}
   \Gamma_{MM\sigma}^M(\vec{k})
   &\to&
   \sum_{i=1,2} \left(g_{\Lambda^* BM}^{(i)} \right)^2 \left( I_{\sigma,M0}^{(i)}I + I_{\sigma,M1}^{(i)} \gamma^0 \right),
   \label{eq:Gamma_EsigM}
   \end{eqnarray}
with
   \begin{eqnarray}
   I_{\sigma,M0}^{(i)}
   &=&
   \frac{g_{MM\sigma}^{(i)}}{(4\pi)^2}
   \int_0^1
   \frac{M_iy}{ h_M^{(i)}(y) }
   dy
   \ ,\\
   I_{\sigma,M1}^{(i)}
   &=&
   \frac{g_{MM\sigma}^{(i)}}{(4\pi)^2}
   \int_0^1
   \frac{M_{\Lambda^*}y^2}{ h_M^{(i)}(y) } dy \ ,
   \end{eqnarray}
where
   \begin{eqnarray}
   h_M^{(i)}(y)
   &=&
   y(m_i^2-M_{\Lambda^*}^2)+y^2M_{\Lambda^*}^2+(1-y)M_i^2\nonumber \\
   &=&
   M_{\Lambda^*}^2y^2+(m_i^2-M_{\Lambda^*}^2-M_i^2)y+M_i^2 \ .
   \label{eq:m^2_M+p^2_M}
   \end{eqnarray}
In this case, both integrations are finite.
   
The $\Lambda^*\Lambda^*\sigma$ interaction Lagrangian should be the same form with Eq.~\ref{eq:Lag_psipsisig}
   \begin{eqnarray}
   {\cal L}^{int}=g_{\Lambda^* \Lambda^* \sigma} \bar{\Lambda^*} \Lambda^* \sigma \ ,
   \end{eqnarray}
which leads to
   \begin{eqnarray}
   \Gamma_{\Lambda^* \Lambda^* \sigma}(\vec{k})
   &=&
   g_{\Lambda^* \Lambda^* \sigma}F_{\Lambda^* \Lambda^* \sigma}(\vec{k}) I
   \label{eq:Gamma_Esig}
   \ .
   \end{eqnarray}
In the leading order of nonrelativistic expansion, we can regard both $I$ and $\gamma^0$ as unity by neglecting the small component of the Dirac spinor.
Combining Eqs.~\ref{eq:Gamma_EsigB}, \ref{eq:Gamma_EsigM} and \ref{eq:Gamma_Esig}, we obtain
   \begin{eqnarray}
   g_{\Lambda^* \Lambda^* \sigma}
   &\sim&
   \frac{1}{F_{\Lambda^* \Lambda^* \sigma}(0)}
   \sum_{i=1,2} \left(g_{\Lambda^* BM}^{(i)} \right)^2 \nonumber \\
   && \times
   \left(
   I_{\sigma,B0}^{(i)} + I_{\sigma,B1}^{(i)} + I_{\sigma,M0}^{(i)} + I_{\sigma,M1}^{(i)}
   \right)
   \label{eq:lamsig_coupling}
   \ .
   \end{eqnarray}
   

\subsection{$\Lambda^*\Lambda^*\omega$ coupling}
Since the $\omega$ meson is a vector meson, the interaction Lagrangian of $\omega$ with two baryon $\psi$ consists of the vector and the tensor terms
   \begin{eqnarray}
   {\cal L}^{int}
   &=&
   g\bar{\psi}\gamma_\mu \psi \omega^\mu
   +
   \frac{f}{4{\cal M}} \bar{\psi}\sigma_{\mu \nu}\psi
   \left(
   \partial^\mu \omega^\nu-\partial^\nu \omega^\mu
   \right)\ ,
   \label{eq:Lag_BBome}
   \end{eqnarray}
with
\begin{eqnarray}
\sigma_{\mu \nu}=\frac{i}{2}[\gamma_\mu,\gamma_\nu]
\ ,
\end{eqnarray}
where ${\cal M}$ is the scaling mass chosen to be the proton mass.
So, the vertices $\tilde{\Gamma}_{BB\omega}$ are given as
   \begin{eqnarray}
   \tilde{\Gamma}_{BB \omega}^{(i)}(\vec{k})
   &=&
   g_{BB \omega}^{(i)}F_{BB\omega}^{(i)}(\vec{k}) \gamma_\mu + i\frac{f_{BB\omega}^{(i)}F_{BB\omega}^{(i)}(\vec{k})}{2{\cal M}}k^\nu\sigma_{\nu\mu}\label{eq:omeBB_vertex} \ ,
   \end{eqnarray}
with $g_{BB \omega}^{(i)}$ ($f_{BB \omega}^{(i)}$) being the vector (tensor) coupling constants of the baryon in channel $i$.
In the $\omega$ meson case, since there exist tensor couplings where the momentum transfer $k^\mu$ contracts with $\gamma_\mu$ vertices, we should take the soft limit after the extraction of the tensor structure. In the zeroth (first) order in $\vec{k}$, the vector (tensor) coupling is given by
   \begin{eqnarray}
   \Gamma_{\Lambda^* \Lambda^* \omega}^B (\vec{k})
   &\to&
   \Gamma_{\Lambda^* \Lambda^* \omega}^B (0) \nonumber \\
   &=&
   \sum_{i=1,2}
   \left(
   g_{\Lambda^* BM}^{(i)}
   \right)^2 \nonumber \\
   && \times
   \left\{ 
   \left(
   I_{\omega g, B0}^{(i)} \gamma_\mu + I_{\omega g, B1}^{(i)} \left\{\gamma_0, \gamma_\mu \right\} + I_{\omega g, B2}^{(i)} \gamma_0\gamma_\mu\gamma_0
   \right)
   \right. \nonumber \\
   &&
   + \frac{i}{2 {\cal M}} \times \nonumber \\
   &&
   \left.
   k^{\nu}
   \Bigl(
    I_{\omega f, B0}^{(i)} \sigma_{\nu \mu}
  + I_{\omega f, B1}^{(i)} \left\{\gamma_0, \sigma_{\nu\mu} \right\}
  + I_{\omega f, B2}^{(i)} \gamma_0 \sigma_{\nu\mu} \gamma_0
   \Bigr)
   \right\} , 
   \end{eqnarray}
with
   \begin{eqnarray}
   I_{\omega g , B0}^{(i)}
   &=&
   \frac{g_{BB\omega}^{(i)}F_{BB\omega}^{(i)}(0)}{(4\pi)^2} \nonumber \\
   &&\times
   \int_0^1 
   y
   \left[
   \frac{M_i^2}{ h_B^{(i)}(y) }
   -\left\{
     a_i(\mu) + \ln \left( \frac{ h_B^{(i)}(y) }{\mu^2} \right)
     \right\}
   \right] dy \ ,\\
   I_{\omega g, B1}^{(i)}
   &=&
   \frac{g_{BB\omega}^{(i)}F_{BB\omega}^{(i)}(0)}{(4\pi)^2}
   \int_0^1
   \frac{M_iM_{\Lambda^*} y(1-y) }{ h_B^{(i)}(y) } dy \ ,\\
   I_{\omega g, B2}^{(i)}
   &=&
   \frac{g_{BB\omega}^{(i)}F_{BB\omega}^{(i)}(0)}{(4\pi)^2}
   \int_0^1
   \frac{M_{\Lambda^*}^2 y(1-y)^2 }{ h_B^{(i)}(y) } dy \ ,\\
   I_{\omega f, B0}^{(i)}
   &=&
   \frac{F_{BB\omega}^{(i)}(0)}{(4\pi)^2}
   \int_0^1 y
   \frac{ f_{BB\omega}^{(i)}M_i^2+2g_{BB\omega}^{(i)}M_i{\cal M} }{ h_B^{(i)}(y) } dy \ ,\\
   I_{\omega f, B1}^{(i)}
   &=&
   \frac{F_{BB\omega}^{(i)}(0)}{(4\pi)^2}
   \int_0^1 
   y(1-y) \frac{f_{BB\omega}^{(i)}M_iM_{\Lambda^*}+g_{BB\omega}^{(i)}M_{\Lambda^*}{\cal M}}{ h_B^{(i)}(y) } dy \ ,\\
   I_{\omega f, B2}^{(i)}
   &=&
   \frac{f_{BB\omega}^{(i)} F_{BB\omega}^{(i)}(0)}{(4\pi)^2}
   \int_0^1
   \frac{ M_{\Lambda^*}^2 y(1-y)^2 }{ h_B^{(i)}(y) } dy \ ,
   \label{eq:Gamma_omeB}
   \end{eqnarray}
with Eq.~\ref{eq:m^2_B+p^2_B}.

Next, we consider the $\omega$ meson and two meson ( $\pi \pi$ or $ K \bar{K} $ ) coupling.
Conservation of G-parity prohibits the $\omega \pi \pi$ coupling, while the $\omega K \bar{K}$ interaction Lagrangian is given by
   \begin{eqnarray}
   {\cal L}_{int}
   &=&
   i g_{\omega K \bar{K}}
   \omega^\mu
   \left[
   \left(
   \partial_{\mu} {\bf K}
   \right)
   \cdot
   \bar{{\bf K}}
   -
   {\bf K}
   \cdot
   \partial_{\mu}
   \bar{{\bf K}}
   \right] .
   \end{eqnarray}
Accordingly we have
   \begin{eqnarray}
   \tilde{\Gamma}^{(1)}_{MM \omega} (\vec{k})
   &=&
   0,
   \label{eq:tilGam_MMome1} \\
   \tilde{\Gamma}^{(2)}_{MM \omega} (\vec{k})
   &=&
   2 g_{\omega K \bar{K}}
   \left(
   \bar{q}_\mu - q_\mu
   \right),
   \label{eq:tilGam_MMome2}
   \end{eqnarray}
where $\bar{q}$ is the average momentum of $\Lambda^*$ in the center of mass frame as
   \begin{eqnarray}
   \bar{q}
   &=&
   \left(
   \sqrt{M_{\Lambda^*}^2 + \frac{\vec{k}^2}{4}}, 0
   \right) ,
   \end{eqnarray}
and $q$ is the variable in the loop integral.
In the same manner as the $\sigma$ meson case, the $\vec{k}$ dependence of the $\omega K \bar{K}$ coupling is not considered.
With the vertices~\ref{eq:tilGam_MMome1} and \ref{eq:tilGam_MMome2}, the vertex $\Gamma_{\Lambda^* \Lambda^* \omega}^M$ is given as a combination of three terms
   \begin{eqnarray}
   \Gamma_{\Lambda^* \Lambda^* \omega}^M (\vec{k})
   &=&
   C_{\gamma_\mu} \gamma_\mu + C_{\bar{q}_\mu} \bar{q}_\mu + C_{k_\mu} k_\mu .
   \end{eqnarray}
The $k_\mu$ term disappears when the soft limit is taken.
For the $\bar{q}$ term, the Gordon identity
   \begin{eqnarray}
   \bar{u}(p') \gamma_{\mu} u(p)
   &=&
   (2M)^{-1}
   \bar{u}(p')
   \left[
   ( p' + p )_{\mu}
   + i \sigma_{\mu \nu}
   ( p'- p)^{\nu}
   \right]
   u(p) ,
   \end{eqnarray}
leads
   \begin{eqnarray}
   \frac{2 \bar{q}_{\mu}}{2M_{\Lambda^*}}
   &=&
   \gamma_{\mu}
   +i \frac{1}{2M_{\Lambda^*}}k^{\nu} \sigma_{\nu \mu} .
   \end{eqnarray}
Therefore, $\Gamma_{\Lambda^* \Lambda^* \omega}^M$ can be rewritten as
   \begin{eqnarray}
   \Gamma_{\Lambda^* \Lambda^* \omega}^M (\vec{k})
   &\to&
   \Gamma_{\Lambda^* \Lambda^* \omega}^M (0) \nonumber \\
   &=&
   \sum_{i=1,2}
   \left(
   g_{\Lambda^* BM}^{(i)}
   \right)^2 \times \nonumber \\
   &&
   \left\{
   \left(
    I^{(i)}_{\omega g, M0} \gamma_\mu
   +I^{(i)}_{\omega g, M1} \left\{ \gamma^0, \gamma_\mu \right\}
   \right) 
   \right.\nonumber \\
   &&
   \left.
   +
   \frac{i}{2 {\cal M}} k^\nu
   \left(
    I^{(i)}_{\omega f, M0}  \sigma_{\nu \mu}
   +I^{(i)}_{\omega f, M1}  \left\{ \gamma^0, \sigma_{\nu \mu} \right\}
   \right)
   \right\} 
   \label{eq:Gamma_omeM},
   \end{eqnarray}
with
   \begin{eqnarray}
   I_{\omega g, M0}^{(2)}
   &=&
   \frac{ g_{MM \omega}^{(2)} }{(4\pi)^2}
   \int_0^1 y
   \left[
   \frac{2(1-y) M_{\Lambda^*} M_2 }{ h_M^{(i)}(y) }
   \right. \nonumber \\
   &&
   \left.
   -
   \left\{
   a(\mu)
   +
   \ln \left( \frac{ h_M^{(i)}(y) }{\mu^2} \right)
   \right\}
   \right] dy,
   \\
   I_{\omega g, M1}^{(2)}
   &=&
   \frac{ g_{MM \omega}^{(2)} }{(4\pi)^2}
   \int_0^1
   \frac{y^2 (1-y) M_{\Lambda^*}^2 }{ h_M^{(i)}(y) }
   dy,
   \\
   I_{\omega f, M0}^{(2)}
   &=&
   \frac{ g_{MM \omega}^{(2)} }{(4\pi)^2}
   \int_0^1
   \frac{2y(1-y) M_2 {\cal M} }{ h_M^{(i)}(y) }
   dy,
   \\
   I_{\omega f, M1}^{(2)}
   &=&
   \frac{ g_{MM \omega}^{(2)} }{(4\pi)^2}
   \int_0^1
   \frac{y^2 (1-y) M_{\Lambda^*} {\cal M} }{ h_M^{(i)}(y) }
   dy ,
   \end{eqnarray}
with Eq.~\ref{eq:m^2_M+p^2_M}.
For the channel $1$, all $I^{(1)}_{\omega}$ are zero .

The $\Lambda^*$ and $\omega$ interaction Lagrangian should take the same form as \ref{eq:Lag_BBome}
   \begin{eqnarray}
   {\cal L}^{int}
   &=&
   g_{\Lambda^* \Lambda^* \omega}
   \bar{\Lambda^* } \gamma_\mu \Lambda^* \omega^\mu
   +
   \frac{f}{4{\cal M}} \bar{\Lambda^*} \sigma_{\mu \nu} \Lambda^*
   \left(
   \partial^\mu \omega^\nu-\partial^\nu \omega^\mu
   \right)\ ,
   \end{eqnarray}
and thus
   \begin{eqnarray}
   \Gamma_{\Lambda^* \Lambda^* \omega}(\vec{k})
   &=&
   g_{\Lambda^* \Lambda^* \omega}F_{\Lambda^* \Lambda^* \omega}(\vec{k})\gamma_\mu + i\frac{f_{\Lambda^* \Lambda^* \omega}F_{\Lambda^* \Lambda^* \omega}(\vec{k})}{2{\cal M}} k^\nu \sigma_{\nu\mu}.
   \end{eqnarray}
Combining Eqs.~\ref{eq:Gamma_omeB} and \ref{eq:Gamma_omeM}, we obtain the $\Lambda^* \Lambda^* \omega$ coupling constants as
   \begin{eqnarray}
   g_{\Lambda^* \Lambda^* \omega}
   &\simeq&
   \frac{1}{F_{\Lambda^* \Lambda^* \omega}(0)}
   \sum_{i=1,2}
   \left(
   g_{\Lambda^* BM}^{(i)}
   \right)^2 \times \nonumber \\
   &&
   \left(
     I_{\omega g, B0}^{(i)} + 2 I_{\omega g, B1}^{(i)} + I_{\omega g, B2}^{(i)}
   + I_{\omega g, M0}^{(i)} + 2 I_{\omega g, M1}^{(i)}
   \right), \\
   f_{\Lambda^* \Lambda^* \omega}
   &\simeq&
   \frac{1}{F_{\Lambda^* \Lambda^* \omega}(0)}
   \sum_{i=1,2}
   \left(
   g_{\Lambda^* BM}^{(i)}
   \right)^2 \times \nonumber \\
   &&
   \left(
     I_{\omega f, B0}^{(i)} + 2 I_{\omega f, B1}^{(i)} + I_{\omega f, B2}^{(i)}
   + I_{\omega f, M0}^{(i)} + 2 I_{\omega f, M1}^{(i)}
   \right),
   \end{eqnarray}
where $\gamma^0$ is regarded as unity.










\end{document}